\documentclass[11pt]{article}

\pdfoutput=1	

\newcommand{\Halmos}{}

\usepackage[colorlinks=true,breaklinks=true,bookmarks=true,urlcolor=blue,citecolor=blue,linkcolor=blue,bookmarksopen=false,draft=false]{hyperref}

\usepackage{geometry}
\usepackage{amssymb}
\usepackage{url}
\usepackage{amsmath}
\usepackage[square,numbers]{natbib}
\usepackage{enumerate}
\usepackage{caption}
\usepackage{bbm}	
\usepackage{epsfig,graphicx}

\usepackage[ruled]{algorithm2e}

\SetAlFnt{\small}
\SetAlCapFnt{\small}
\SetAlCapNameFnt{\small}
\SetAlCapHSkip{0pt}
\IncMargin{-\parindent}

\newtheorem{theorem}{Theorem}[section]

\newtheorem{example}{Example}[section]

\newtheorem{lemma}[theorem]{Lemma}
\newtheorem{proposition}[theorem]{Proposition}

\newtheorem{claim}[theorem]{Claim}
\newtheorem{definition}[theorem]{Definition}

\newtheorem{problem}{Open Problem}

\def\squarebox#1{\hbox to #1{\hfill\vbox to #1{\vfill}}}
\newcommand{\qed}{\hspace*{\fill}
	\vbox{\hrule\hbox{\vrule\squarebox{.667em}\vrule}\hrule}\smallskip}

\newenvironment{proof}{\noindent{\bf Proof:~~}}{\(\qed\)}

\usepackage{csquotes}

\newcommand{\EF}{\operatorname{EF}}
\newcommand{\EFstar}{\operatorname{EF-1^*}}
\newcommand{\PO}{\operatorname{\widetilde{PO}}}
\newcommand{\EFone}{\operatorname{EF-1}}
\newcommand{\alloc}{\mathcal{S}}
\newcommand{\allocA}{\mathcal{A}}
\newcommand{\allocB}{\mathcal{B}}
\newcommand{\allocC}{\mathcal{C}}

\newcommand{\hatone}{\hat{\alloc}^1}
\newcommand{\hattwo}{\hat{\alloc}^2}

\begin{document}

\title{Competitive Equilibrium with Indivisible Goods and Generic Budgets} 

\author{Moshe Babaioff\thanks{Microsoft Research, {\tt moshe@microsoft.com}.}, Noam Nisan\thanks{The Hebrew University of Jerusalem and Microsoft Research, {\tt noam@cs.huji.ac.il}.}, Inbal Talgam-Cohen\thanks{Technion -- Israel Institute of Technology, {\tt italgam@cs.technion.ac.il}.}}

\date{}

\maketitle

\begin{abstract}
Competitive equilibrium from equal incomes (CEEI) is a classic solution to the problem of fair and efficient allocation of goods [Foley'67, Varian'74]. Every agent receives an equal budget of artificial currency with which to purchase goods, and prices match demand and supply. However, a CEEI is not guaranteed to exist when the goods are indivisible, even in the simple two-agent, single-item market. Yet, it is easy to see that once the two budgets are slightly perturbed (made \emph{generic}), a competitive equilibrium does exist. 

In this paper we aim to extend this approach beyond the single-item case, and study the existence of equilibria in markets with two agents and additive preferences over multiple items. We show that for agents with \emph{equal} budgets, making the budgets \emph{generic} -- by adding vanishingly small random perturbations -- ensures the existence of an equilibrium. We further consider agents with \emph{arbitrary} non-equal budgets, representing \emph{non}-equal entitlements for goods. We show that competitive equilibrium guarantees a new notion of fairness among non-equal agents, and that it exists in cases of interest (like when the agents have identical preferences) if budgets are perturbed. Our results open opportunities for future research on generic equilibrium existence and fair treatment of non-equals.

\end{abstract}
		
\maketitle

\section{Introduction.}
\label{sec:intro}
We study fair and efficient allocation of indivisible goods in settings in which utility is non-transferable (without money).
Such settings arise in many real-life allocation scenarios, such as allocating donated food items to food banks \cite{Pre17}, assigning courses to students or shifts to workers~\cite{Bud11}, or sharing scientific/computational resources within a university or company. 
Our motivation is two-fold: First, we wish to identify conditions for the existence of a competitive equilibrium despite the goods' indivisibility. 
Second, we wish to develop notions of fairness that apply to the allocation of such goods among agents with different entitlements; we approach fairness through the prism of competitive equilibrium -- a connection made long ago by \cite{Fol67,Var74}.

\subsection{Competitive equilibrium in Fisher markets with indivisible goods.}
A fundamental achievement of general equilibrium theory is to establish conditions under which a competitive equilibrium exists in markets of divisible items \citep[Chapters 15--20]{AD54, Eis61, MWG95}). The theory does not extend to indivisible items. 
In this paper we focus on a simple Fisher market model \citep{BS05}, where there are $m$ goods and $2$ agents with positive budgets $b_1,b_2$ of artificial currency, and each agent has an additive preference over the goods. 
Unfortunately, even in the simplest markets of this structure, 
when there is a  single indivisible item and two agents with the same budget ($m=1$ and $b_1=b_2=1/2$), 
no competitive equilibrium exists. This is so since if the item's price is at most $1/2$ then both agents demand it, while if its price is above $1/2$ then neither agent can afford it and the market does not clear.\footnote{Since the currency is artificial and money has no intrinsic value, an agent is not satisfied if the price is exactly her budget but she doesn't get the item.}

The non-existence of a competitive equilibrium in the above simple example is a knife-edge phenomenon: if the budgets were $1/2+\epsilon$ and $1/2$, instead of being \emph{exactly} equal, then an equilibrium would exist (e.g., by setting the item price at $1/2+\epsilon$.) 
To avoid knife-edge non-existence, \citet{Bud11} initiated the study of competitive equilibrium with \emph{almost-equal} budgets.
He showed existence results for an \emph{approximate} such equilibrium notion, in which the demand and supply are only approximately matched by the prices.  
In this paper we take the idea of almost-equal budgets a step further. We study competitive equilibrium with {\em generic} budgets -- arbitrary budgets (possibly far from equal) that are slightly perturbed. 
Unlike Budish, we focus on monotone and in particular additive preferences and thus are able to avoid approximate equilibrium notions. 

Our main conceptual contribution is pointing out that making the budgets generic significantly expands the set of markets in which competitive equilibria exist. 
In this paper we demonstrate the promise of our approach of perturbing the budgets (making them generic) for obtaining provable existence guarantees in several settings of interest, including the case of two agents with equal budgets (which are perturbed to become generic) or different budgets but the \emph{same} additive preferences.  
Moreover, evidence gathered from computer simulations (and a small amount of real-life data -- see Appendix \ref{appx:computer}) suggests that generic equilibrium existence may be quite common in practical settings. 
This motivates the study of competitive equilibrium from generic budgets as a solution concept for indivisible items.
Nevertheless, we note that generic budgets are not a silver bullet solution to competitive equilibrium non-existence in every market, as we show in our companion paper \cite{BNT18} for agents with non-additive preferences.\footnote{In particular, for 2 agents and 5 items, when the agents' preferences are monotone yet \emph{non-additive}, generic budgets are not sufficient to guarantee existence of competitive equilibria.}
We note that once equal budgets are perturbed, or when agents have \emph{non}-equal entitlements for the goods (represented by non-equal budgets) standard notions of fairness (like proportional share) are no longer appropriate. We thus turn next to discuss fairness properties with non-equal entitlements.

\subsection{Fairness with indivisible goods and different entitlements.}  

Which allocations of goods among agents should be considered ``fair''? A vast literature is devoted to this question (for expositions see, e.g., 
\citep[Chapters 11--13]{BT96,BCE+16}). Many standard notions of fairness in the literature reflect an underlying assumption that all agents have an equally-strong claim to the goods. However, some agents may be \emph{a priori} much more entitled to the goods than others (see, e.g. \citep[Chapters 3 and 11]{RW98}). Real-life examples include -- in addition to the ones mentioned above -- partners who own different shares of the partnership's holdings~\citep{CGK87}, different departments sharing a company's computational resources \citep{GZH+11}, or family members splitting an heirloom \citep{PZ90}. 
Our model captures these situations, since the entitlement of agent~$i$ can be encoded by his budget $b_i$ (we may assume without loss of generality that $\sum_i b_i = 1$).
 
An appropriate notion of fairness should of course take into account the indivisible goods and different entitlements, and capture the idea of ``proportional satisfaction of claims'' \cite[][p.~95]{Bro90}.
In the special case where all entitlements are equal ($b_1=...=b_n=1/n$),
\emph{envy-freeness} (no agent prefers another's allocation to his own) is an important fairness criterion \citep{Fol67}, but it is not clear how to generalize this notion to heterogeneous entitlements. 
\emph{Fair share} (every agent  $i$ prefers his allocation to a $b_i$-fraction of all items, when divisible) is also an important fairness criterion \citep{Ste48}, but it does not easily extend to indivisible items. 

In this paper we adapt a well-known approach to fair allocation with equal entitlements -- finding a \emph{competitive equilibrium from equal incomes} (\emph{CEEI}) \citep{Var74,BM16} -- to the case of unequal entitlements. We may treat the agents as buyers and their entitlements as budgets, and seek a competitive equilibrium in the resulting Fisher market. 
Such an equilibrium is not guaranteed to exist, but this is to be expected -- indivisibility can indeed undermine the ability to fairly allocate. 
Our idea is to first perturb the entitlements (budgets), 
and aim to show that competitive equilibrium from such \emph{generic} incomes
exists more widely and offers approximate fairness guarantees that are the best possible given indivisibilities. 
Thus, a better understanding of competitive equilibrium with generic budgets can help reach fair division of indivisible items among agents with heterogeneous entitlements.

\subsection{Our results.}

In this paper we focus on markets in which agents have additive preferences. 
This is a natural starting point for research on generic budgets, which already raises rich technical challenges and requires novel techniques.
Additive preferences have been the main focus of recent research on fair allocation of indivisible goods \cite{BM16}.
Moreover, we focus on the case of 2 agents, which is a natural starting point and is probably the most important in practice (e.g., the fair-division website Fair Outcomes Inc.~is dedicated to 2 agents, implementing the 2-agent adjusted winner protocol \cite{BT96}). See \cite[Section 1.1.1]{PR17} for more on the importance of the 2-agent setting.

As we saw, for two agents with the same budgets and a single item, a competitive equilibrium (CE) fails to exist. 
In that example, each agents' proportional share of the item  is half the item, yet one necessarily is getting nothing. 
We first prove that in any case that there is a way to avoid this unfair situation, a CE indeed exists, and this is true even when budgets are not equal. 
Define the \emph{budget-proportional share} of agent $i$ to be any set of items that that gives the agent at least $b_i/(b_1+b_2)$ of her value for the set $M$ (her total value).
We prove that if there is an allocation that concurrently gives each of the two agents her budget-proportional share, then a CE exists. 
In the single item case and agents with equal budgets, clearly there is no allocation that gives both agents their proportional share. We thus proceed to our main contribution -- using budget perturbations and by that moving to generic budgets -- as a tool to obtain existence of equilibrium even when there is no allocation that gives the agents their budget-proportional shares, and this is obtained even when there are multiple items.  

\subsubsection{Main existence result: Generic budgets circumvent CEEI nonexistence.}
Our main result is that for two additive agents, even with multiple items, generic budgets can serve as a form of tie-breaking to solve the CEEI nonexistence issue in our setting: 

\begin{theorem}[Existence for almost equal budgets]
	\label{thm:intro-main1}
	Consider 2 agents with additive preferences over multiple items. 
	If the budgets are sufficiently close to being equal, but not precisely so, 
	then a CE exists.
\end{theorem}

As we have seen,  for a single item any such Ce cannot give both agents their proportional share. Yet, our proof shows that the CE that is guaranteed to exist by Theorem \ref{thm:intro-main1}, gives each agent her \emph{truncated share}. The truncated share is defined by her share at the agent's most preferred Pareto efficient allocation, assuming she is doomed to get less than her proportional share. 

We leave open the problem of extending the theorem to more than 2 agents. 

\begin{problem}
	Consider $n>2$ agents with additive preferences over multiple items. 
	Does there exist a CE when budgets are sufficiently close to being equal?	
\end{problem}

\subsubsection{Main fairness result for competitive equilibria.}

For agents with equal budgets, it is known that when a CE exists, its allocation
has a natural fairness property related to the well-known cut-and-choose protocol. In this protocol one agent (the ``cutter'') divides the items into 2 sets, and the other agent chooses her favorite set. Each agent gets a bundle she prefers at least as much as the one she can guarantee for herself as the cutter -- this is called her \emph{maximin share (MMS)}. 

When budgets are perturbed, they are no longer equal. Thus we would like to understand the fairness properties guaranteed when budgets are not equal.  
How can the above protocol be adapted to different budgets? 
We next suggest a generalization of MMS to the case of different budgets:

An \emph{$\ell$-out-of-$d$ maximin share} of an agent is any bundle at least as preferred as the one she can guarantee by the following (hypothetical) protocol: the agent partitions the items into $d$ parts (some may be empty), and then takes the worst $\ell$ of these parts (since the other agents are entitled to $d-\ell$ of the parts, and their choice may as well be the worst possible for the cutter).  

\begin{theorem}[Fairness]
	\label{thm:intro-fairness} 
	Consider $n$ agents with general preferences. In any CE allocation, for every agent $i\in [n]$ and rational number $\ell/d\le b_i$, agent $i$ gets her $\ell$-out-of-$d$ maximin share.
\end{theorem}

\subsubsection{Additional results.}
Theorem \ref{thm:intro-main1} is based on a characterization we develop for CEs with 2 agents and arbitrary budgets, which has a particularly nice form for additive preferences. From the characterization it follows that linear combinations of the agents' preferences form equilibrium prices under a certain condition (``budget exhaustion''). We use a graphical representation of allocations and their values (depicted in Figures \ref{fig:main-proof-prelim}-\ref{fig:equal}) to identify allocations and value combinations for which the condition holds, and CE existence follows. 

We use the machinery we have built to prove some additional interesting results for markets with two additive agents, presented below. 
 
{\it Second welfare theorem.}
The  characterization we develop for CEs with 2 agents is also useful in establishing the second welfare theorem in our setting. Despite the fundamental nature of the theorem and setting, to the best of our knowledge this was not previously observed.
The theorem states that for any valuations, any Pareto efficient allocation for these valuations is an allocation of some CE, for some appropriately set budgets. 

\begin{theorem}[Second welfare theorem]
	\label{thm:intro-second}
	Consider 2 agents with additive preferences.
	For every Pareto efficient allocation $\alloc = (\alloc_1, \alloc_2)$, there
	exist budgets $b_1, b_2$ and prices $p$ such that $(\alloc,p)$ is a CE.
\end{theorem}

{\it Competitive equilibrium with non-equal budgets.}
The same machinery that establishes existence for almost equal budgets applies to agents with very different budgets, 
to which the solution concept of CE with generic budgets immediately extends. 
We focus on the case in which the agents have the same preferences, putting them in the most direct competition over the goods, and for this case we show: 

\begin{theorem}[Generic existence for identical preferences, different budgets]
	\label{thm:intro-different}
	Consider 2 agents with identical additive preferences over the goods and arbitrary budgets.
	If their budgets are generic then a CE exists.	
\end{theorem}

\subsection{Organization.}

Section \ref{sec:model} introduces the model and preliminaries.
Section \ref{sec:fairness} addresses the fairness properties of CEs (formally states and proves Theorem \ref{thm:intro-fairness}).
Section \ref{sec:characterization} begins to build the tools we will need for our existence results, by characterizing CEs with 2 agents (and deriving Theorem \ref{thm:intro-second} as a consequence).
Section \ref{sec:proportional} shows that if both agents can receive their budget-proportional share then a CE exists. 
The required machinery for existence with generic budgets is set up in Section \ref{sec:technical}.
Our main result (Theorem \ref{thm:intro-main1}) follows, and is formally stated and proved in Section~\ref{sec:almost-equal-budgets}.
Section \ref{sec:different-budgets} includes the result for different budgets (Theorem \ref{thm:intro-different}).
Section \ref{sec:discussion} contains a discussion and conclusions.

\subsection{Additional related work.}

\paragraph{Discrete Fisher markets.}
Our model is a special case of Arrow-Debreu exchange economies with indivisible items. Several other variants have been studied:
\citep{Sve83,Mas87,ADG91,FY02,NW16} consider markets with indivisibilities in which an infinitely divisible good plays the role of money, and so money carries inherent value for the agents. Shapley and Scarf 
\citep{SS74}, Svensson \citep{Sve84} and subsequent works focus on the house allocation problem with \emph{unit-demand} agents. Several works assume a continuum of agents \citep[e.g.,][]{Mas77}, and/or study relaxed CE notions \citep[e.g.,][]{Die71,DPS03,RC07}. Closest are the models of \emph{combinatorial assignment} \cite{Bud11}, which allows non-monotonic preferences, and of \emph{linear markets} 
\cite{DPS03}, which crucially relies on non-generic budgets.

\paragraph{CEEI.}
\citet{Bud11} circumvents the non-existence of CEEI due to indivisibilities by weakening the equilibrium concept and allowing market clearance to hold only approximately. He focuses exclusively on budgets that are almost equal. In the same model, \citet{OPR16} show PPAD-completeness of computing an approximate CEEI, and NP-completeness of deciding the existence of an approximate CEEI with better approximation factors than those shown to exist by Budish. The preferences used in the hardness proofs are non-monotone, leading Othman et al.~to suggest the research direction of restricting the preferences (as we do here) as a way around their negative results. \citet{BHM15} study (exact) CEEI existence for two valuation classes (perfect substitutes and complements) with non-generic budgets.
For \emph{divisible} items, there has been renewed interest in CEEI under additive preferences due to their succinctness and practicality. \citet{BM16} offer a characterization based on natural axioms, and \citet{BMSY16} analyze CEEI allocations of ``bads'' rather than goods.

\paragraph{Fairness with indivisibilities.} 
Most notions of fair allocation in the classic literature apply to divisible items. 
\citet{BCM16} survey research on fairness with \emph{indivisible} items, emphasizing computational challenges (see also \citep{dKB+09,AGMW15}). The works of \citep{LMMS04,MP11}
study envy minimization (rather than elimination) with indivisibilities. 
\citet{GHK+05} study envy-freeness in the context of profit-maximizing pricing. 
Several recent works \cite[][]{CG15,CKM+16,BGM16,CDG+17,AMOV18,GHM18} consider the approximation of Nash social welfare and related fairness guarantees. 
As for practical implementations of fair division with indivisibilities, these are discussed by \citet{BC12} (for responsive preferences), \citet{OSB10} (implementing findings of \citet{Bud11}), \citet{GP14} (presenting the Spliddit website for additive preferences), and \citet{BT96} (presenting the adjusted winner algorithm in which one item may need to be divided). 
Mechanism design aspects appear in~\cite{ABCM17}.
For further discussion see Section \ref{sec:fair}.

\paragraph{Concurrent and subsequent work.} 
Several additional results related to CEs with generic budgets appear in our companion paper~\cite{BNT18}, which studies preferences that are not necessarily additive. The paper considers the case of general monotone preferences and maps the limits of this approach for different numbers of items. 
It then considers cardinal preferences over many items and defines a hierarchy of preference classes, establishes relations between them and proves results about existence of equilibria with generic budgets for some of these classes. 

In a follow-up paper to our work, \cite{Seg18} has shown that for four agents with arbitrary budgets, non-existence persists even with generic budgets. Yet, that result does not rule out that for more than 2 agents, equilibrium exists for generic budgets that are almost equal.

Recently there has been increasing interest in fair division of indivisible goods; the following works appeared concurrently or subsequently to early versions of our paper:

\citet{FGH+17} independently develop a new fairness notion for allocation among agents with cardinal preferences and different entitlements. 
Their notion is distinct from ours and is not directly related to the solution concept of a CE.
For example, their fairness notion is not always guaranteed when allocating 3~items among agents with generic budgets (as implied in their Theorem 2.1). This is at odds with our CE existence results and resulting fairness guarantees according to our notions of fairness (Proposition~\ref{pro:gen-maximin} and companion paper), demonstrating the difference between the approaches. 

Several recent papers have advanced our understanding of envy-freeness, including \cite{PR18} which studies the envy-free relaxation $\EFstar$ suggested by \cite{CKM+16}, and the works of \cite{ABC+18} and \cite{CS18} which study envy-freeness with incomplete knowledge.

\section{Model.}
\label{sec:model}
In this section we formulate our market model and competitive equilibrium notion (with generic budgets), and present fairness preliminaries.

Throughout we use the notation $[d]$ where $d$ is a positive integer to denote the set $\{1,\dots,d\}$.

\subsection{Market setting.}

We study discrete Fisher markets which consist of a set $M$ of  $m$ indivisible items and a set $N$ of $n$ agents.
We refer to subsets of items as bundles and often denote them by $S$ or $T$.
Each agent $i\in N$ has a \emph{cardinal} preference, represented by a valuation function $v_i:2^M\to \mathbb{R}_{\ge 0}$ which assigns to every bundle $S$ a nonnegative value~$v_i(S)$. Agent $i$ prefers bundle $S$ to bundle $T$ iff $v_i(S) > v_i(T)$,
and cardinality allows us to compare by how much -- by considering the ratio $v_i(S)/v_i(T)$. The absolute values themselves however don't matter, hence a cardinal valuation can be normalized without loss of generality (wlog). 

Our results hold primarily for the class of \emph{additive} preferences. A cardinal preference $v_i$ is additive if $v_i(S)=\sum_{j \in S}v_i(\{j\})$ for every bundle~$S$. 
We assume wlog that $v_i$ is normalized, i.e., 
$$
v_i(M)=1.
$$
We also assume that preferences are \emph{monotone} (satisfy free disposal), i.e., $v_i(S) < v_i(T)$ whenever $S \subset T$. Moreover we assume \emph{strict} preferences (no indifferences), i.e., either $v_i(S) < v_i(T)$ or $v_i(S) > v_i(T)$ whenever $S\ne T$. We allow one exception to strictness -- items may be identical, in which case preferences over them are allowed to be weakly-monotone rather than strictly so, reflecting that identical items are interchangeable.%
\footnote{No agent wants multiple copies of the same item. All the CEs in our existence results assign the same prices to identical items. This is similar to the treatment of identical items in \cite{Bud11}, where prices are assigned to classes rather than to individual seats.}

In addition to preferences, agents in our model have \emph{budgets}. Let $b=(b_1,b_2,\ldots,b_n)$ be a budget profile, where $b_i>0$ is agent $i$'s budget. 
Unless stated otherwise, we assume wlog that 
$$
\sum_{i=1}^{n} b_i=1,
$$ 
and $b_1\geq b_2\geq \ldots \geq b_n$ (every market can be converted to satisfy these properties by renaming and normalization).

We emphasize that budgets and ``money'' in our model are only means for allocating items among agents with different entitlements. Money has no intrinsic value for the agents, whose preferences are over subsets of items and who disregard any leftover budget. For this reason, it is important that the preferences are scale-free, in contrast to a model with money where a value $v_{i,j}$ can be interpreted as the amount agent $i$ is willing to pay for item $j$. Similarly, scale-dependent measures of social efficiency and fairness like welfare and minimum-value are inappropriate in our model. 

\subsection{Competitive equilibrium (CE).}

The goal of the market is to allocate the items among the agents.
An \emph{allocation} $\alloc=(\alloc_1,\alloc_2,\dots,\alloc_n)$ is a 
partition of \emph{all} items among the agents, i.e., $\alloc_i \cap \alloc_k = \emptyset$ for $i \ne k$ and $\bigcup_i \alloc_i =M$. 
By definition, an allocation is \emph{feasible} (no item is allocated more than once) and \emph{market clearing} (every item is allocated).%

A competitive equilibrium is an allocation together with item prices that ``stabilize'' it.
Let $p = (p_1,p_2,\ldots, p_m)$ denote a vector of non-negative item prices. The price $p(S)$ of a bundle $S$ is then $\sum_{j\in S} p_j$. We say that $S$ is \emph{within budget~$b_i$} if $p(S)\leq b_i$, and that it is \emph{demanded} by agent $i$ at price vector $p$ if it is the most preferred bundle within her budget at these prices.
Formally, $p(\alloc_i) \le b_i$, and $p(T) > b_i$ for every $T$ such that $v(T)> v(\alloc_i)$. We can now define our equilibrium notion:

\begin{definition}[CE]
	A \emph{competitive equilibrium (CE)} is a pair $(\alloc,p)$ 
	of allocation $\alloc$ and item prices $p$, such that $\alloc_i$ is demanded by agent $i$ at prices $p$ for every $i\in N$.
\end{definition}

Given a market with preferences $\{v_i\}_{i\in N}$ and budget profile~$b$, an allocation $\alloc$ is \emph{supported} in a CE if there exist item prices $p$ such that $(\alloc,p)$ is a CE. 
Where only preferences are given, we overload this notion and say that $\alloc$ is supported in a CE if there exist prices $p$ and budgets $b$ such that $(\alloc,p)$ is a CE.

Given budgets $b$ and an allocation $\alloc$, we say that prices $p$ are \emph{budget-exhausting} if $p(\alloc_i)=b_i$ for every agent $i$. Note that if $p$ is budget-exhausting then every agent is allocated, i.e., $\alloc_i\neq \emptyset$ for every $i$ (since $b_i>0$). 
We observe that budget-exhaustion is wlog when every agent is allocated, and use this observation throughout the paper. 

\begin{claim}[Budget-exhausting prices are wlog]
	\label{cla:exhaust-wlog}
	For every CE $(\alloc,p)$ such that $S_i\neq \emptyset$ for every $i\in N$, 
	there exists a CE $(\alloc,p')$ in which $p'$ is budget-exhausting.
\end{claim}

\proof{Proof.}
	As every agent is allocated at least one item, we can raise the price of that item in his allocated bundle until her budget is exhausted. The new prices form a CE with the original allocation since every agent still gets her demanded set. 
	\Halmos
\endproof

\subsubsection{Pareto optimality (PO).}

What does it means for an allocation (CE or otherwise) to be ``efficient''?

\begin{definition}[PO]
	Consider a market with preferences $\{v_i\}_{i\in N}$. An allocation $\alloc$ is \emph{Pareto optimal (PO)} (a.k.a.~Pareto efficient) if no allocation $\alloc'$ \emph{dominates} $\alloc$, 
	i.e., if for every $\alloc'\ne\alloc$ there exists an agent~$i$ for whom $v_i(\alloc_i) > v_i(\alloc'_i)$.
\end{definition}

We use the notation $\PO=\PO(v_1,v_2)$ to denote the set of all different PO allocations for preferences $v_1,v_2$.

There are two kinds of fundamental welfare theorems in economics that apply to various market equilibrium notions. The first is about Pareto optimality and holds for CEs in our setting; we discuss the second in Section \ref{sec:characterization}.

\begin{theorem}[First welfare theorem]
	\label{thm:first-welfare-thm}
	Let $(\alloc,p)$ be a CE. Then $\alloc$ is PO.
\end{theorem}

\proof{Proof (for completeness).} 
Assume for contradiction an alternative allocation $\alloc'$, such that for every agent~$i$ for whom $\alloc_i\ne \alloc'_i$ it holds that $v_i(\alloc_i)< v_i(\alloc'_i)$. Consider the total payment $\sum_i p(\alloc'_i)$ for the alternative allocation given the CE prices $p$. By market clearance, $\sum_i p(\alloc'_i)= \sum_i p(\alloc_i)$. Therefore there must exist an agent $i$ for whom $\alloc_i\ne \alloc'_i$ but $p(\alloc'_i)\le p(\alloc_i)$. This means that $\alloc_i$ cannot be demanded by agent $i$, leading to a contradiction.
\Halmos
\endproof

\subsubsection{Generic budgets (vs.~different budgets).}
\label{sub:generic-def}

We are interested in showing \emph{generic} existence of a CE for classes of markets. 
We use the standard notion of genericity, i.e., ``all except for a zero-measure'', or equivalently, ``with tiny random perturbations''. By generic existence we thus mean that for every market in the class, for every vector of budgets except for a zero-measure subset, a CE exists. 
An equivalent way to say this is: for every market in the class, for every vector of budgets, by adding tiny random perturbations to the budgets we get a new instance in which a CE exists with probability 1.

A useful way of specifying a zero-measure subset of budgets is as those which satisfy some condition, for instance, $b_1=2b_2$. The conditions we use differ among different CE existence results, and for concreteness we shall list them explicitly within each result; we emphasize however that the conditions themselves are irrelevant to our contribution, as long as the measure of budgets satisfying them is zero.

Generic budgets are not to be confused with \emph{different} (or \emph{arbitrary}) budgets, by which we mean budgets that are not necessarily equal or almost equal to one another. While fair allocation among (almost) equal agents is well-studied, much less is known for agents who have a priori different entitlements, as modeled by different budgets.

\subsection{Fairness preliminaries.}
\label{sec:fair} 

We include here the fairness preliminaries most related to our results; a more detailed account -- including \emph{ordinal} preferences in addition to cardinal ones, envy-freeness in addition to fair share, and Nash social welfare -- appears in Appendix \ref{sec:fair} and is summarized in Tables \ref{tab:one}-\ref{tab:two}. These tables also show where our new fairness notions fit in with existing ones. 

The ``two most important tests of equity'' according to \citet[][p.166]{Mou95} are (i) guaranteeing each agent his \emph{fair share (FS)}; and (ii) \emph{envy-freeness (EF)}. Our main concern is FS, ``probably the least controversial fairness requirement in the literature'' \citep{BMSY16} -- in the setting of indivisible items and agents with different budgets for which not much is known. 
Intuitively, FS for \emph{divisible} items and \emph{equal-budget} agents guarantees that each agent believes she receives at least $1/n$ of the divisible ``cake'' (while EF guarantees she believes no one else receives a better slice than hers). More formally,
FS requires for each agent to receive a bundle that she prefers at least as much as the bundle consisting of a $1/n$-fraction of every divisible item on the market. 

When items are \emph{indivisible}, to define FS we must use the cardinal nature of the preferences in our model.
The parallel of FS is the notion of proportionality, which extends naturally to agents with different budgets: Given a budget profile~$b$, an allocation $\alloc$ gives agent $i$ his \emph{budget-proportional share} if agent $i$ receives at least a $b_i$-fraction of his value for all items, that is 
$v_i(\alloc_i)\geq b_i \cdot v_i(M)$.
An allocation is \emph{budget-proportional} (a.k.a.~weighted-proportional) if every agent receives his proportional share. When all budgets are equal, such an allocation is simply called \emph{proportional}. 

Unfortunately, budget-proportionality is a very restrictive fairness requirement when dealing with indivisible items (see Section \ref{sec:proportional}). \citet{Bud11} studies the following weaker notion:

\begin{definition}[FS with indivisibilities]
	\label{def:maximin-share}
	An allocation $\alloc$ guarantees \emph{1-out-of-$n$ maximin share} if every agent receives a bundle she prefers at least as much as the bundle she can guarantee for herself by the following procedure: partitioning the items into $n$ parts, and allowing the $n-1$ other agents to chose their parts first (assuming their choice is the worst possible for her). 
\end{definition}

Every CE for~$n$ agents with \emph{equal} budgets gives every agent his 1-out-of-$n$ maximin share and achieves EF. An allocation that gives every agent his 1-out-of-$n$ maximin share does not always exist beyond 2 agents \cite{KPW18}.
\citet{Bud11} shows that every CE for~$n$ agents with \emph{almost} equal budgets guarantees 1-out-of-$(n+1)$ maximin share (as if there were an extra agent to share with). In Proposition \ref{pro:gen-maximin} we generalize this result to \emph{different} budgets, by defining the fairness notion of $\ell$-out-of-$d$ maximin share.


\section{Fairness Properties of CE.}
\label{sec:fairness}
Fairness properties that apply when agents have equal entitlements are no longer appropriate even when agents have almost-equal budgets. Indeed, when there is only 1 item, one of the agents is allocated nothing and thus does not get his proportional share.
Thus we are interested in fairness properties that are guaranteed by the existence of a CE when agents have non-equal budgets (possibly even far from being equal). In a sense, we are building upon the classic connection between CE with \emph{equal} budgets and fairness, expanding it to \emph{different} budgets, and using it to derive a natural fairness notion appropriate for a priori non-equal agents. 

In Section \ref{sub:l-out-of-d-def} we define our notion (Def.~\ref{def:l-out-of-d}): a parameterized version of the $1$-out-of-$n$ maximin share guarantee (Def.~\ref{def:maximin-share}), generalizing it to accommodate arbitrary (possibly very different) budgets. 
In Section \ref{sub:l-out-of-d-exist} we show that every CE guarantees fairness according to our notion (Proposition \ref{pro:gen-maximin}), 
and in Section \ref{sub:l-out-of-d-dicuss} we briefly discuss implications. 
(We remark that in subsequent sections we discuss a different fairness notion -- truncated share -- which is guaranteed by all of our CE existence results but not by every CE in general.)

\subsection{Definition of $\ell$-out-of-$d$ maximin share.}
\label{sub:l-out-of-d-def}

Consider an agent $i$. Her $\ell$-out-of-$d$ maximin bundle is the bundle she can guarantee for herself by the following (hypothetical) protocol: the agent partitions the items into $d$ parts, lets the other agents choose $d-\ell$ of these parts (their choice is assumed to be worst-case), then receives the remaining $\ell$ parts. 

\begin{example}
The 2-out-of-3 maximin bundle of an additive agent who values items $(A,B,C)$ at $(1,2,3)$ is $\{A,B\}$. 
\end{example}

Agent $i$'s $\ell$-out-of-$d$ maximin share reflects how preferable her $\ell$-out-of-$d$ maximin bundle is for her.
We now give a formal definition:

\begin{definition}
	\label{def:l-out-of-d}
	An allocation $\alloc$ guarantees agent~$i$ her \emph{$\ell$-out-of-$d$ maximin share} if 
	$$
	v_i(\alloc_i) \ge \max_{\text{partition } (T_1,\dots,T_d)} \left\{\min_{L\subseteq[d],|L|=\ell} \left\{v_i\left(\bigcup_{t\in L} T_{t}\right) \right\} \right\}.
	$$
\end{definition}

\subsection{$\ell$-out-of-$d$ maximin share in equilibrium.}
\label{sub:l-out-of-d-exist}

The following proposition holds generally for any market setting (for any number of agents, preference class, etc.), and shows that a CE guarantees for every agent her fair share in the sense of Definition \ref{def:l-out-of-d}. The parameters $\ell$ and $d$ in the $\ell$-out-of-$d$ maximin share guarantee correspond to the budget $b_i$ of the agent, thus mirroring her a priori entitlement to the items.

\begin{proposition}
	\label{pro:gen-maximin}
	Let $b$ be an arbitrary budget profile. Every CE guarantees agent $i$ her $\ell$-out-of-$d$ maximin share for every rational number $\ell/d\le b_i$.
\end{proposition}

\proof{Proof.}
Let $(S,p)$ be a CE and let $P$ denote the sum of prices $\sum_{j\in M} p_j$. Since $S$ is an allocation of all items, every item is ``purchased'' by an agent and so $P = \sum_{j} p_j\leq \sum_{i}b_i = 1$.
Let $(T_1,\ldots,T_d)$ be any partition of the items into $d$ parts, and observe that  
$1\geq P=\sum_j p_j= p(\bigcup_{t=1}^d T_t)= \sum_{t=1}^d p(T_t)$ (using linearity of the prices). 
By the pigeonhole principle, there exists a subset of $\ell$ parts whose total price is at most ${\ell\over d}P  \leq  {\ell\over d}$. Let us call this ``the cheap subset''. By the proposition's assumption, ${\ell\over d}\leq b_i$. Therefore, agent $i$ can afford the cheap subset, and by definition of a CE the bundle actually allocated to agent $i$ must be at least as preferred by him as the cheap subset.
\Halmos
\endproof

\subsection{Discussion of $\ell$-out-of-$d$ maximin share.}
\label{sub:l-out-of-d-dicuss}


For the case of almost equal budgets, Proposition \ref{pro:gen-maximin} subsumes the celebrated result of \citet{Bud11} that every CE with almost equal budgets gives every agent her $1$-out-of-$(n+1)$ maximin share. To see this notice that if budgets are almost equal then $b_1\ge b_2 \ge \dots \ge b_n \ge \frac{n}{n+1}b_1\ge \frac{1}{n+1}$ for every agent~$i$ (the last inequality follows since $b_1$ must be $\ge 1/n$ for the budgets to sum up to~1). Thus the result of Budish can be deduced from Proposition \ref{pro:gen-maximin}. 

Proposition~\ref{pro:gen-maximin} also strengthens the result of Budish in the following sense. Consider $n=2$ a priori equal agents and $m=5$ items. The $1$-out-of-$(n+1)$ maximin share in this case guarantees the worst part out of a partition of the $5$ items into $3$ parts; the $\ell$-out-of-$d$ maximin share applies with $\ell=2,d=5$ (since $\ell/d=2/5\ll 1/2$), guaranteeing the worst $2$ parts out of a partition of the $5$ items into $5$ parts. If an agent views the items as roughly equal, the latter guarantee is strictly better ($2$ items rather than $1$).

The flexibility allowed by the parameters $\ell$ and $d$ is even more important when dealing with different (i.e.~not almost equal) budgets.
Consider $m=3$~items and an agent with budget $5/13$. 
Dividing the 3 items into $13$ parts and taking the worst $5$ parts does not guarantee the agent anything beyond an empty bundle, whereas taking the worst part among $3$ (using that $1/3\le 5/13$) guarantees her at least one item.

\section{CE Characterization and Second Welfare Theorem.}
\label{sec:characterization}
In this section we state and prove a sufficient condition for CE existence (Lemma \ref{lem:budget-exhaust}), which is the workhorse of our existence results in subsequent sections. We demonstrate its usefulness by deriving from it the second welfare theorem (Theorem \ref{thm:second-welfare}).

\subsection{Characterization.}
\label{sub:charac}

We begin by presenting necessary and sufficient conditions for a budget-exhausting pricing and PO allocation to form a CE, when each of the $n=2$ agents gets a non-empty set (in this case budget-exhaustion is wlog by Claim~\ref{cla:exhaust-wlog}). The characterization holds beyond additive preferences for any pair of cardinal preferences.
To state it we use the following standard notation: for a preference $v_i$ and disjoint sets $S,T$, the \emph{marginal value} of $S$ given $T$ is denoted by $v_i(S\mid T) = v_i(S\cup T) - v_i(T)$.

\begin{proposition}[Characterization]
	\label{pro:char}
	Given 2 agents with monotone cardinal preferences $v_1,v_2$, consider a PO allocation $\mathcal S=(\alloc_1,\alloc_2)$ in which $\alloc_i\ne\emptyset$ for $i\in \{1,2\}$, and budget-exhausting item prices $p$.
	Then $(\mathcal S,p)$ forms a CE if and only if for  $i,k\in \{1,2\}, i\neq k$, and for every two bundles $S\subseteq \alloc_i,T\subseteq \alloc_{k}$, 
	\begin{equation}
	v_i(S\mid \alloc_i\setminus S)>v_i(T\mid \alloc_i\setminus S) \text{~~and~~} v_{k}(S\mid \alloc_{k}\setminus T)>v_{k}(T\mid \alloc_{k}\setminus T) \implies p(S)>p(T).\label{eq:cond}
	\end{equation} 
\end{proposition}

\proof{Proof.}
For the first direction, assume by way of contradiction that $(\mathcal S,p)$ is a CE but Condition~(\ref{eq:cond}) is violated. 
Wlog assume that this is the case for $S\subseteq \alloc_1$ and $T\subseteq \alloc_2$, i.e., it holds that 
$v_1(S\mid \alloc_1\setminus S)>v_1(T\mid \alloc_1\setminus S) \text{~~and~~} v_{2}(S\mid \alloc_{2}\setminus T)>v_{2}(T\mid \alloc_{2}\setminus T)$ while
$p(S)\le p(T)$. 
Then agent 2 prefers to swap $T$ for $S$ and has enough budget to do so, in contradiction to the fact that he gets his demanded set in the CE.

For the other direction, consider a pair $(\mathcal S,p)$ such that Condition~(\ref{eq:cond}) holds for every $S,T$ as in the proposition statement. 
Assume by way of contradiction that (wlog) agent 2 is not allocated his demanded set. Then since agent 2's budget is exhausted, this means there must be bundles $S\subseteq \alloc_1$ and $T\subseteq \alloc_2$ such that $v_2(S\mid \alloc_2\setminus T)>v_2(T\mid \alloc_2\setminus T)$ and $p(S)\le p(T)$. 
Therefore, $v_1(S\mid \alloc_1\setminus S)\leq v_1(T\mid \alloc_1\setminus S)$, 
and so by swapping $S,T$ in the allocation we arrive at a new allocation strictly preferred agent $2$, and no worse for agent $1$, 
in contradiction to the Pareto optimality of~$\alloc$.
\Halmos
\endproof

\subsection{Sufficient condition for equilibrium existence.}
\label{sub:sufficient}

When preferences are additive, prices can be derived from weighted linear combinations of the preferences:

\begin{definition}
	\label{def:comb-price}
	Consider 2 additive preferences $v_1,v_2$, and parameters $\alpha, \beta\in\Re_+$ such that $\max\{\alpha,\beta\}>0$.
	The \emph{combination pricing} $p$ with parameters $\alpha,\beta$ is an item pricing that assigns every item $j$ the price $p_j=\alpha v_{1}(\{j\})+\beta v_{2}(\{j\})$. 
\end{definition}

Observe that by additivity of $v_1,v_2$ in Definition \ref{def:comb-price}, the combination pricing $p$ with parameters $\alpha,\beta$ assigns every bundle $S$ the price $p(S)=\alpha v_1(S)+\beta v_{2}(S)$. 
Note that identical items have identical prices (items $j,j'$ are identical precisely if $v_i(\{j\})= v_i(\{j'\})$ for every agent $i$). 

The following lemma presents a sufficient condition for CE existence, and is the workhorse of our existence results in this paper:

\begin{lemma}[Budget-exhausting combination pricing is sufficient]
	\label{lem:budget-exhaust}
	Consider 2 agents with additive preferences and budgets $b_1\ge b_2>0$ (possibly equal). If for a PO allocation $\alloc$ there exists a budget-exhausting combination pricing $p$, then $(\mathcal S, p)$ is a CE.
\end{lemma}

\proof{Proof.}
The existence of a budget-exhausting pricing indicates that both agents are allocated nonempty bundles in $\mathcal S$. Thus by Proposition \ref{pro:char}, to prove the lemma it is sufficient to show that Condition~\eqref{eq:cond} holds. For additive preferences this condition can be written as: for every $i,k\in \{1,2\},i\ne k$, and for every two bundles $S\subseteq \alloc_i,T\subseteq \alloc_{k}$, 
\begin{equation*}
v_i(S)>v_i(T) \text{~~and~~} v_{k}(S)>v_{k}(T) \implies p(S)>p(T).\label{eq:simple-cond}
\end{equation*} 	
Plugging in the combination pricing, for every $S,T$ such that $v_1(S)>v_1(T)$ and $v_2(S)>v_2(T)$, it holds that $p(S)=\alpha v_1(S)+\beta v_2(S)>\alpha v_1(T)+\beta v_2(T)=p(T)$. So Condition \eqref{eq:cond} holds for any $i\neq k\in \{1,2\}$, and $(\mathcal S,p)$ is a CE.
\Halmos
\endproof

\subsection{Second welfare theorem.}
\label{sub:second}

A second fundamental theorem of welfare economics is of the form~\cite[][Part III, p.~308]{MWG95}:
\begin{displayquote}
	``{\it [A]ny Pareto optimal outcome can be achieved as a competitive equilibrium if appropriate lump-sum transfers of wealth are arranged}.'' 
\end{displayquote}
In particular, this means that any socially-efficient allocation that a social planner deems desirable for its equitability can be realized in equilibrium.
In our context, such a theorem would say that for every set of agents and their preferences, for every PO allocation~$\alloc$ of items among them, we can find budgets for the agents and prices for the items which support $\alloc$ as a CE.
From the companion paper we know that such a theorem does \emph{not} hold for 2 agents with general preferences. We use Lemma \ref{lem:budget-exhaust} to establish the second welfare theorem for two agents with  additive preferences:

\begin{theorem}[Second welfare theorem]
	\label{thm:second-welfare}
	Consider 2 agents with additive preferences. For every PO allocation $\mathcal S$, there exist budgets $b_1,b_2$ and prices $p$ for which $(\mathcal S,p)$ is a CE. 
	
	Moreover, if both agents are allocated non-empty bundles in $\alloc$, then $(\mathcal S,p)$ is a CE for \emph{any} combination pricing $p$ and corresponding budgets $b_1=p(\alloc_1), b_2=p(\alloc_2)$.
\end{theorem}

\proof{Proof.}
If $\alloc$ allocates all items to a single agent, wlog~agent 1, then we can get a CE by pricing every item at some arbitrary price $\rho>0$, setting $b_1=m\rho$, and setting $b_2<\rho$ (the budgets can of course be normalized). 
Otherwise, fix any combination pricing $p$; in particular, a combination pricing with parameters $\alpha=\beta=1$. Set $b_i=p(\alloc_i)$ for every agent $i$. By Lemma~\ref{lem:budget-exhaust}, $(\alloc,p)$ is a CE, completing the proof.
\Halmos
\endproof

\section{Budget-Proportionality.}
\label{sec:proportional}
Section \ref{sub:prop-exist} shows that if we are fortunate enough to face a setting in which a budget-proportional allocation exists despite item indivisibility, then a CE is also guaranteed to exist (Theorem \ref{thm:prop2CE}). This existence result does \emph{not} rely on generic budgets. Section~\ref{sub:truncated} deals with the case of no budget-proportional allocation, identifying two alternative candidates for CE allocations which are as close to budget-proportional as possible. These are the basis for the existence results in Sections \ref{sec:technical}-\ref{sec:different-budgets}.

\subsection{Equilibrium existence when a budget-proportional allocation exists.}
\label{sub:prop-exist}

Since preferences are normalized, agent $i$ gets his budget-proportional share precisely when $v_i(\alloc_i)\geq b_i$. 
We say that agent $i$ gets \emph{at most} his budget-proportional share if $v_i(\alloc_i)\le b_i$, and call an allocation \emph{anti-proportional} if every agent gets at most his budget-proportional share and at least one agent gets strictly below it
(Claim~\ref{obs:anti-proportional} in Appendix~\ref{appx:fairness} demonstrates the existence of markets in which the allocation of every CE is anti-proportional.)

We show that every budget-proportional PO allocation is supported in a CE. By the same argument, anti-proportional PO allocations are also supported in a CE. The proof is by exploiting (anti-)proportionality to construct a budget-exhausting combination pricing.
 
\begin{proposition}
	\label{prop:CE-budget-prop}
 	For 2 agents with additive preferences and \emph{any} budgets, every budget-proportional PO allocation is supported in a CE.
 	%
 	Additionally, every anti-proportional PO allocation is supported in a CE. 
\end{proposition}

\proof{Proof.}
Let $v_1, v_2$ be the preferences and $b_1,b_2$ the budgets (all normalized), and let $\alloc=(\alloc_1,\alloc_2)$ be a budget-proportional PO allocation.
 
Since $v_1(\alloc_1)\geq b_1$, $v_2(\alloc_2)\geq b_2$ and by normalization,
we have that $v_1(\alloc_2) = 1-v_1(\alloc_1)\leq 1-b_1= b_2$ and $v_2(\alloc_1) = 1-v_2(\alloc_2)\leq 1-b_2= b_1$.

We now construct a budget-exhausting combination pricing with parameters $\alpha,\beta$.
If it is the case that $v_1(\alloc_1) + v_2(\alloc_2)=1$ (and thus $b_1=v_1(\alloc_1)$ and  $b_2=v_2(\alloc_2)$), we set $\alpha=1$ and $\beta=0$.
Otherwise, $v_1(\alloc_1) + v_2(\alloc_2)-1\neq 0$, and we can set 
\begin{equation}
	\alpha = \frac{v_2(\alloc_2)-b_2}{v_1(\alloc_1) + v_2(\alloc_2)-1}, \beta = 1-\alpha = \frac{v_1(\alloc_1)-b_1}{v_1(\alloc_1) + v_2(\alloc_2)-1}.\label{eq:alpha-beta}
\end{equation}
Observe that 
$$
\alpha v_1(\alloc_1) + \beta v_2(\alloc_1) = \frac{v_2(\alloc_2)-b_2}{v_1(\alloc_1) + v_2(\alloc_2)-1} v_1(\alloc_1) + \frac{v_1(\alloc_1)-b_1}{v_1(\alloc_1) + v_2(\alloc_2)-1} (1-v_2(\alloc_2)) = b_1,
$$ 
and that similarly $\alpha v_1(\alloc_2) + \beta v_2(\alloc_2) = b_2$. Since the allocation is budget-proportional, it holds that $\alpha,\beta\geq 0$.
In each of the two cases we thus have a combination pricing $p$ with parameters $\alpha,\beta$ such that $p(\alloc_1)=b_1$ and $p(\alloc_2)=b_2$, and thus by Lemma \ref{lem:budget-exhaust},  $(\mathcal S, p)$ is a CE. 	

It remains to consider anti-proportional PO allocations. For every such allocation in which one agent gets at most his budget-proportional share and the other gets less then his share, the same pair of parameters $\alpha,\beta$ defined in Equation \eqref{eq:alpha-beta} will give a budget-exhausting combination pricing with non-negative parameters, and thus a CE.
\Halmos
\endproof

Observe that every budget-proportional allocation is dominated by a budget-proportional PO allocation. The following theorem thus follows directly from Proposition \ref{prop:CE-budget-prop}.

\begin{theorem}
	\label{thm:prop2CE}
	If there exists a budget-proportional allocation then a CE exists.
\end{theorem}

\subsection{As close as possible to budget-proportionality.}
\label{sub:truncated}

Let us assume from now on that a budget-proportional allocation does not exist (and nor does an anti-proportional PO allocation). 
When it is not possible to simultaneously give each agent her budget-proportional share, the next best thing in terms of fairness is her ``truncated'' share: the best she can obtain in any PO allocation in which she is allocated \emph{at most} her budget-proportional share. In this section we formalize this notion, and show two PO allocations that give both agents their truncated shares. Both of these are natural candidates for CEs, and we indeed validate this intuition in Sections \ref{sec:technical}-\ref{sec:different-budgets}. Thus, all of our positive results establish the existence of CEs for allocations that are ``as fair as possible''. In Example \ref{example:CE-no-trunc-share} we show that not every CE has this property. 

\begin{figure}[htb]
	\begin{center}
	\includegraphics[width=5in,height=3in]{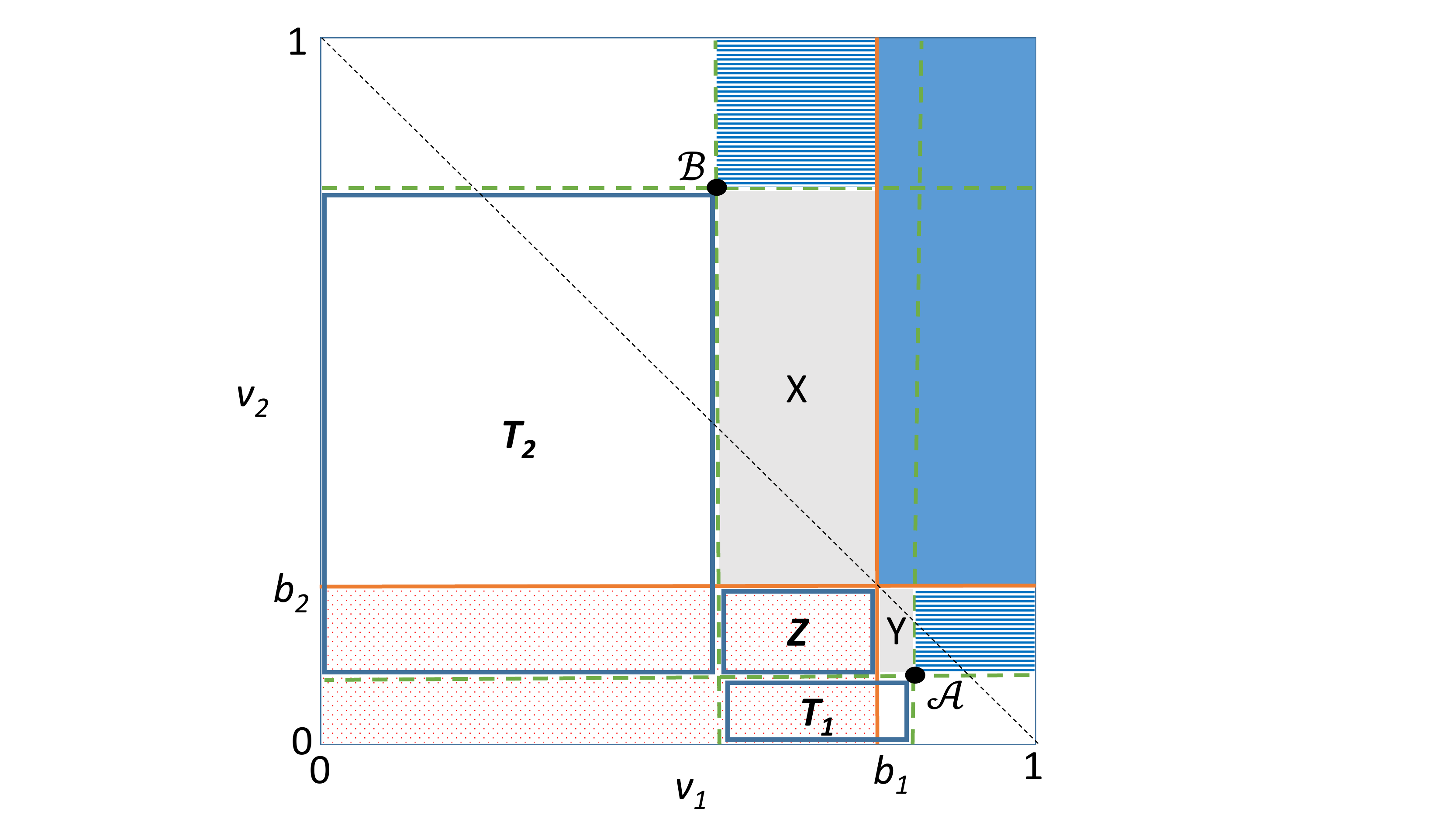}
	\end{center}
	\caption{This figure illustrates the setting and notation when neither a budget-proportional nor an anti-proportional PO allocation exists. It shows the value of an allocation for agent 1 on the $v_1$-axis and the value of an allocation for agent 2 on the $v_2$-axis. Every allocation $\alloc$ can be represented by the point $(v_1(\alloc_1), v_2(\alloc_2))$ on the $(v_1\times v_2)$-plane. 
	The two points $\allocA,\allocB\in \PO$ represent two PO allocations.	
	The agents' budgets $b_1,b_2$ are shown on the same axes. 
	\newline
	The closure of the solid blue area (at or to the right of $b_1$ and at or above $b_2$) includes all allocations that are budget-proportional, and is empty by assumption. The closure of the red dotted area (at or to the left of $b_1$ and at or below $b_2$) represents anti-proportional allocations and has no PO allocations by assumption. 
	By Pareto optimality, the blue striped areas (to the right and above $\allocA$ and $\allocB$) are both empty -- the only allocation in their closures are $\allocA$ and $\allocB$. The figure also depicts rectangles $T_1,T_2,X,Y$, and $Z$, which will play a role in our arguments in Sections \ref{sec:technical}-\ref{sec:different-budgets}.}
	\label{fig:main-proof-prelim}
\end{figure}

Figure~\ref{fig:main-proof-prelim} depicts the setting and fixes our notation. 

\begin{definition}[\bf Truncated share]
	\label{def:trunc-share}
	Let $b_i^-=\max_{\alloc\in \PO \mid v_i(\alloc_i)\leq b_i} \{v_i(\alloc_i)\}$ be the maximum share agent~$i$ can obtain in any PO allocation in which she gets at most her budget-proportional share. Denote by $\hat{\alloc}^i=\hat{\alloc}^i(b_i)$ 
	the maximizing PO allocation, i.e., $b_i^-=v_i(\hat{\alloc}^i_i)$.%
	\footnote{The allocation  $\hat{\alloc}^i(b_i)$ is well-defined: it is possible to give nothing to agent $i$, so the maximum is taken over a non-empty set of allocations. It is unique up to identical items.}
	An allocation $\alloc$ gives agent $i$ her truncated budget-proportional share, or \emph{truncated share} for short, if $v_i(\alloc_i)\geq b_i^-$.
\end{definition}

An analogous definition is the following:

\begin{definition}[\bf Augmented share]
	\label{def:aug-share}
	Let $b_i^+=\min_{\alloc\in \PO \mid v_i(\alloc_i)\geq b_i} \{v_i(\alloc_i)\}$ be the minimum share agent~$i$ can obtain in any PO allocation in which she gets at least her budget-proportional share. Denote by $\check{\alloc}^i= \check{\alloc}^i(b_i)$ 
	the minimizing PO allocation, i.e., $b_i^+=v_i(\check{\alloc}^i_i)$.
	An allocation $\alloc$ gives agent~$i$ her \emph{augmented share} (and thus in particular her truncated share) if $v_i(\alloc_i)\geq b_i^+$.
\end{definition}

The following lemma establishes a simple but important fact about the four allocations $\hat{\alloc}^i,\hat{\alloc}^k,\check{\alloc}^i,\check{\alloc}^k$, which give agents $i,k$ their truncated or augmented shares, respectively. Namely, it turns out that these four allocations are in fact two, since $\hat{\alloc}^i=\check{\alloc}^k$ and $\hat{\alloc}^k=\check{\alloc}^i$. By definition, each of these two allocations gives each agent at least her truncated share.

\begin{lemma}[Two ``as fair as possible'' allocations]
	\label{lem:same-alloc}
	Consider 2 agents $i\ne k\in\{1,2\}$ with additive preferences and arbitrary budgets. 
	Assume there are no budget-proportional allocations nor PO anti-proportional allocations.  
	Then the PO allocation $\check{\alloc}^i$ coincides with the PO allocation $\hat{\alloc}^k$. 
	That is, $\check{\alloc}^i$ obtains share $b_i^+$ for agent $i$ and share $b_k^-$ for agent $k$. 
\end{lemma}

\begin{figure}
	\begin{center}
	\includegraphics[width=5in,height=3in]{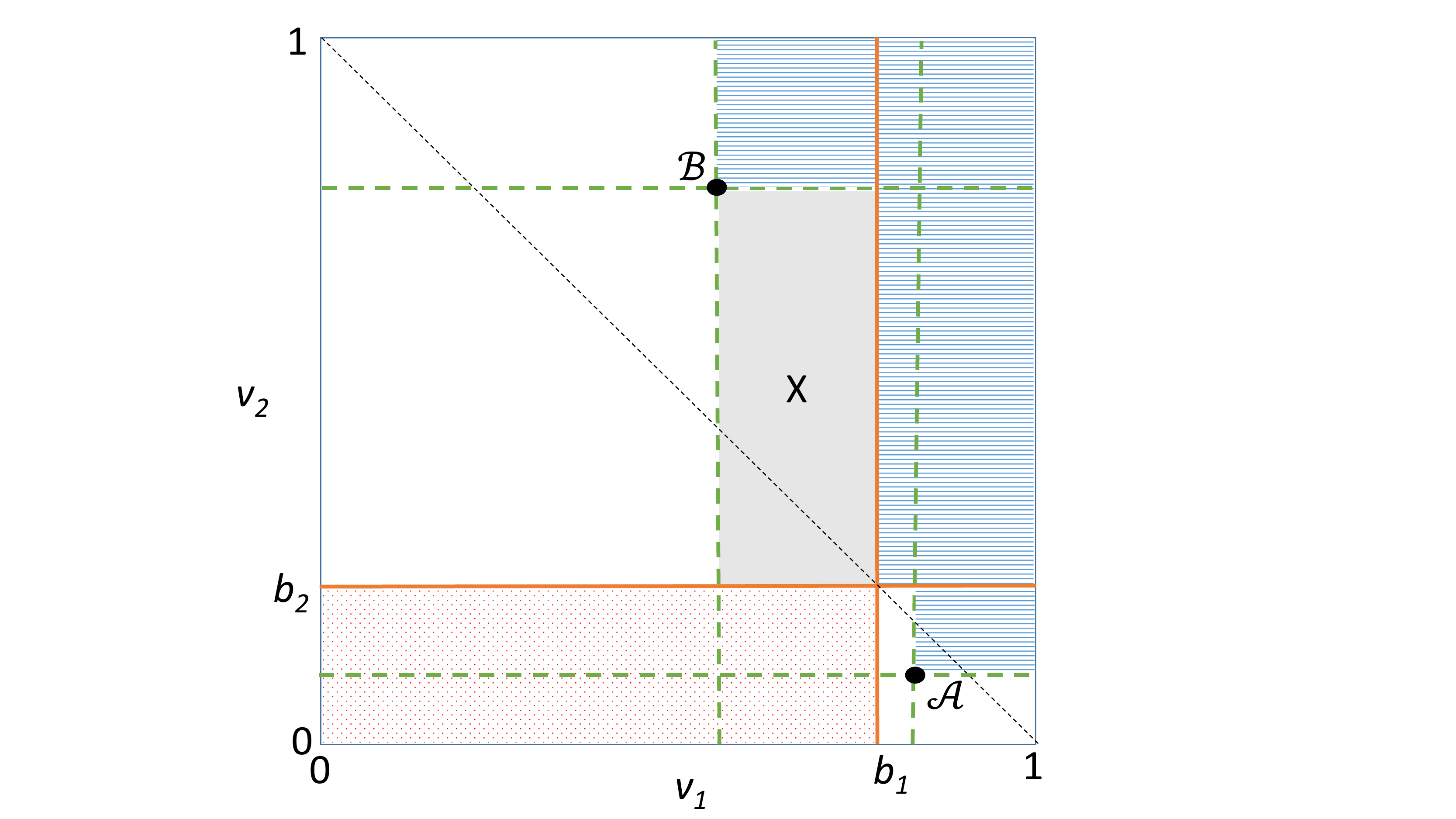}
	\end{center}
	\caption{Illustration of the proof of Lemma \ref{lem:same-alloc}.}
	\label{fig:2-points-not-4}
\end{figure}

\proof{Proof.}
	Let $k=1, i=2$ (the complementary case $k=2, i=1$ is similar). Denote $\allocA=\check{\alloc}^1$ and $\allocB=\check{\alloc}^2$. 
	The notation we use for the proof is depicted in Figure \ref{fig:2-points-not-4}, where indeed in allocation $\allocA$ agent 1 can be seen to receive value above $b_1$, and in allocation $\allocB$ agent 2 can be seen to receive value above $b_2$. We now use Figure \ref{fig:2-points-not-4} to argue that $\allocB=\hatone$ (showing that $\allocA=\hattwo$ is similar).
	
	By assumption, the closure of the blue striped area (union of the areas to the right and above $\allocA$ and $\allocB$, with the area at or to the right of $b_1$ and at or above $b_2$) is empty of allocations except for $\allocA$ and $\allocB$ (the same was established for Figure~\ref{fig:main-proof-prelim}). 
	By definition, $\allocB$ is the lowest PO allocation at or above $b_2$. Any PO allocation at or to the left of $b_1$ that is to the right of $\allocB$ must be in the interior of the solid gray rectangle $X$ (using that there are no PO anti-proportional allocations, i.e., no PO allocations in the closure of the dotted red area). Yet such a point in the interior of $X$ is not only to the right of  $\allocB$, it is also below it. This means that it is closer than $\allocB$ to $b_2$ from above, yielding a contradiction. 
	
	The closed rectangle $X$ must therefore be empty of PO allocations except for $\allocB$ (and thus must also be empty of non-PO allocations). But the point corresponding to allocation $\hatone$ must fall within the closure of rectangle $X$ by definition (it is the rightmost PO point at or to the left of $b_1$, and it cannot lie to the left of $\allocB$). Thus $\hatone=\allocB$, completing the proof.
	\Halmos
\endproof		
	
The following example demonstrates that not every CE gives every agent her truncated share.

\begin{example}[Less fair than possible]
	\label{example:CE-no-trunc-share}
	Consider 2 additive agents who both value items $(A,B,C,D)$ at $(7.9, 1, 5, 2)$ (unnormalized). The budgets are $b_1=\frac{1}{2}+\epsilon,b_2=\frac{1}{2}-\epsilon$ for some sufficiently small~$\epsilon$. Every allocation is PO. The allocation $(\{B,C,D\},\{A\})$ is an equilibrium allocation in which agent 2 gets a share of $\frac{7.9}{15.9}<b_2$. The allocation $(\{A,B\},\{C,D\})$ is also an equilibrium allocation, despite the fact that agent 2's share drops to $\frac{7}{15.9}$.%
	\footnote{The supporting prices for the two allocations are, respectively, $p=(\frac{1}{2}-\epsilon,\frac{1}{6},\frac{1}{6},\frac{1}{6}+\epsilon)$ and $p=(\frac{1+\epsilon}{2},\frac{\epsilon}{2},\frac{1}{4},\frac{1}{4}-\epsilon)$.}
\end{example}

\section{Main Technical Tool.}
\label{sec:technical}
In this section we build our main technical tool for establishing generic existence of CEs for agents with almost equal budgets (Section \ref{sec:almost-equal-budgets}), and for agents with the same preferences and different budgets (Section \ref{sec:different-budgets}).
This tool is formalized in Lemma \ref{lem:2-CE-exist-suff}, which establishes two conditions that together are sufficient for a CE to exist. Namely, for a fixed agent $i$, the conditions are:
\begin{enumerate}
	\item Genericity of the budgets, defined as \emph{not} belonging to a zero-measure subset $R_i$; 
	\item Emptiness of ``rectangle'' $T_i$ from any allocations (see Figure \ref{fig:main-proof-prelim}).
\end{enumerate}
The genericity condition is what drives our existence results, and is therefore to be expected.
The condition that $T_i$ is empty, however, is a necessary artifact of our proof techniques.%
\footnote{Our simulations identified an example in which for each agent $i$, $T_i$ is not empty, and there is no CE with item prices based on scaling $v_i(\{j\})$ (other CEs were found).
The example includes~7 items. Agent 1's values are $(0.1420, 0.0808, 0.1921, 0.1717, 0.1651, 0.1200, 0.1283)$, agent 2's values are $(0.0827, 0.1056, 0.1743, 0.1515, 0.1862, 0.1123, 0.1874)$. The budgets are $0.8093$ and $0.1907$.
\cite{GK18} constructed a similar example with~8 items in which there is no CE with \emph{combination} pricing (other CEs were found). 
}
Dropping this condition requires novel ideas, and we leave this as an open question for future research.

\subsection{Definitions.} 

We now formally define $R_i,T_i$. For the definition of $T_i$, recall the allocations $\hat{\alloc}^i,\hat{\alloc}^k$ (Definition \ref{def:trunc-share}).

\begin{definition}[Rectangle of allocations $T_i$]
	\label{def:T-i}
	Let $T_i=T_i(b_i,v_1,v_2)$ be the set of allocations $\alloc$ satisfying $v_i(\hat{\alloc}^i_i)< v_i(\alloc_i) < v_i(\hat{\alloc}^k_i)$  and $0 < v_k(\alloc_k)< v_k(\hat{\alloc}^k_k)$. 
\end{definition}

For the definition of $R_i$, let $d=|\PO|$ be the total number of PO allocations. Order all allocations in $\PO$ by agent $i$'s preference, such that his $r$-th least preferred PO allocation is at index $r\le d$. Denote this allocation by $\alloc(r)$, so that $v_i(\alloc(r+1)_i) > v_i(\alloc(r)_i)$ for every index $r\le d-1$. 

\begin{definition}[Zero-measure subset of budgets $R_i$]
	\label{def:R-i}
	Every budget pair $(b_i,1-b_i)$ for agents $(i,k)$, respectively, belongs to $R_i=R_i(v_1,v_2)$ iff there exists an index $r$ such that $\frac{b_i}{v_i(\alloc(r+1)_i)}= \frac{1-b_i}{1 - v_i(\alloc(r)_i)}$. 
\end{definition}

Note that 
$R_i$ is a zero-measure subset of the budget pairs. 

\subsection{Statement and proof.}

\begin{lemma}[Main technical tool.]
	\label{lem:2-CE-exist-suff}
	Consider 2 agents with additive preferences $v_1,v_2$ and budgets $b_1>b_2$. 
	Assume there are no budget-proportional allocations nor PO anti-proportional allocations.
	If for some agent~$i$, $(b_1,b_2)\notin R_i(v_1,v_2)$ and the set $T_i= T_i(b_i,v_1,v_2)$ is empty, then a CE exists. 	
	Moreover, in this CE every agent gets his truncated share. 
\end{lemma}

\proof{Proof.}
Let $i$ be the agent for which the conditions of the lemma hold. By Lemma~\ref{lem:same-alloc}, both PO allocations $\check{\alloc}^1$ and $\check{\alloc}^2$ give both agents their truncated share. To prove the claim it is thus sufficient to show that at least one of these allocations is supported in a CE. We next show that indeed for some $\gamma_i\in (0,1)$, at least one of these two allocations is supported by item prices of the form $p_j=\gamma_i v_i(\{j\})$ for every item $j$.

\begin{figure}[htb]
	\begin{center}
	\includegraphics[width=5in,height=3in]{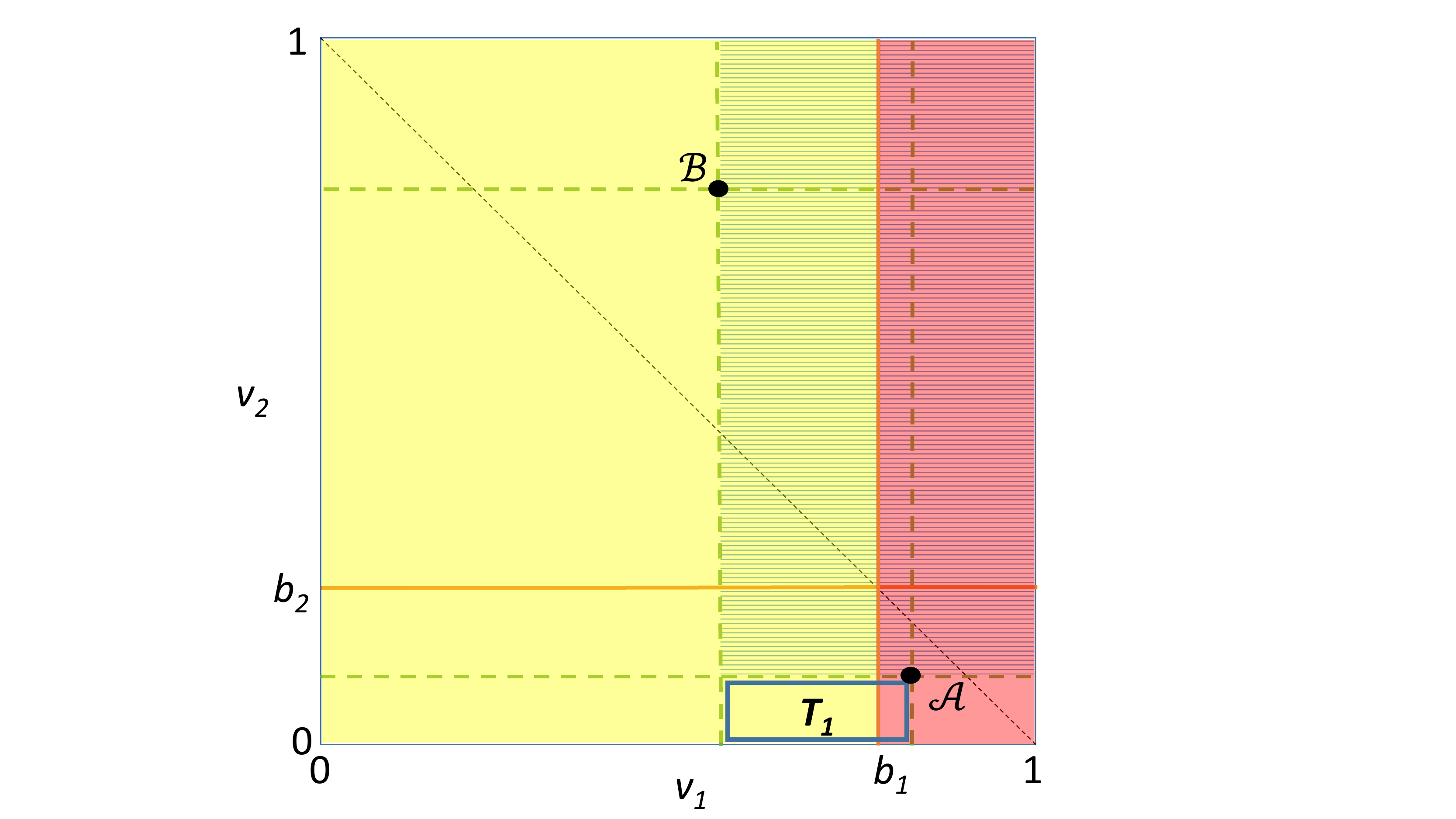}
	\end{center}
	\caption{This figure illustrates the first part of the proof of Lemma \ref{lem:2-CE-exist-suff}, using the notation of Figure~\ref{fig:main-proof-prelim}. It shows the allocations that each agent can afford given his budget when prices are $p=v_1$ (i.e., according to agent 1's valuation). Allocations in the solid yellow rectangle (at or to the left of $b_1$) have value at most  $b_1$ for agent 1, and thus also price at most $b_1$, so agent $1$ can afford them. Allocations in the solid red rectangle (at or to the right of $b_1$) have value at least $b_1$ for agent 1, and thus agent 1 values agent 2's allocation at most at $1-b_1=b_2$ (by normalization), so the price is at most $b_2$ and affordable for agent~$2$. 
	\newline	
	The blue striped area marks allocations with value for agent $1$ that is above his value for $\allocB$, and value for agent $2$ that is above his value for $\allocA$. This area has no allocation at all, as it is subset of the union of the following areas: the blue areas from Figure \ref{fig:main-proof-prelim} without allocations below $\allocB$ and to the left of $\allocA$; the interiors of $X$ and $Y$ that are empty by the proof of Lemma \ref{lem:same-alloc} in Figure \ref{fig:2-points-not-4}; and the interior of $Z$ that must be empty, as an allocation there must be dominated by some PO allocation in the areas we just argued are empty, or by an anti-proportional PO allocation (which does not exist by assumption). 
	\newline	
	Therefore, if rectangle $T_1$ is empty then at these prices agent $1$ demands the allocation $\allocB= \check{\alloc}^2$ (the rightmost allocation within the yellow area -- his budget), while agent~$2$ demands the allocation $\allocA= \check{\alloc}^1$ (the highest allocation within the red area -- his budget).\label{fig:gamma1}}
	{}
\end{figure}

We first characterize the set of allocations that are within the budget of each agent when prices are set to $p_j=\gamma_i v_i(\{j\})$ for every item $j$, and $\gamma_i \in (0,1)$ (i.e., prices are a linearly scaled down version of agent $i$'s valuation). Agent $i$ can afford any allocation $\alloc$ such that $\gamma_i v_i(\alloc_i)\leq b_i$. Agent $k$ can afford any allocation $\alloc$ such that  $\gamma_i v_i(\alloc_k)\leq b_k $, or equivalently $\gamma_i (1-v_i(\alloc_i))\leq 1-b_i$ (using that both valuations and budgets are normalized, that is, $b_1+b_2=v_1(M)=v_2(M)$). We illustrate this for $\gamma_i=1$ in Figure \ref{fig:gamma1}. 

\begin{figure}[htb]
	\begin{center}
	\includegraphics[width=5in,height=3in]{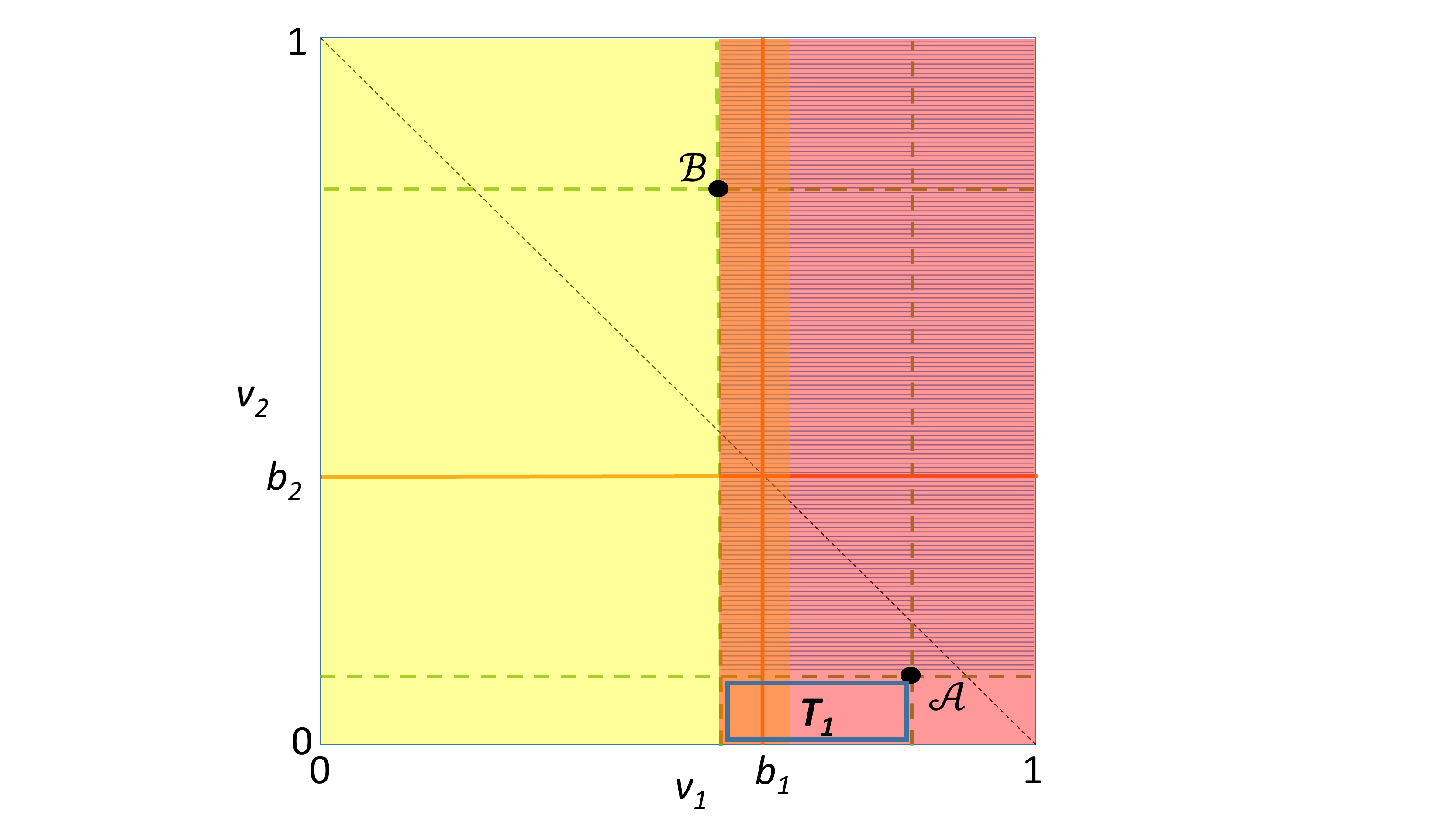}
	\end{center}
	\caption{This figure illustrates Case 2 in the proof of Lemma \ref{lem:2-CE-exist-suff}, using the notation of Figure~\ref{fig:gamma1}, for agent $i=1$. Prices are $p_j=\gamma_1 v_1(\{j\})$ for every item $j$ where $\gamma_1 =b_2/v_1(\check{\alloc}^2_2)$. 
	The solid yellow and red rectangles (starting strictly to the right of $b_1$ and stretching to the left, and at or to the right of $\allocB$, respectively) are the allocations that agents 1 and 2 can afford, respectively. Both agents can afford more allocations than when $\gamma_i=1$ (cf.~Figure \ref{fig:gamma1}). The overlap (depicted as an orange rectangle) contains the allocations that both agents can afford. 
	\newline
	The value of $\gamma_1$ is such that agent 2 can exactly afford allocation $\allocB$, which is clearly demanded by him at these prices (the highest allocation within his budget, using that the blue striped area is empty and that $\allocB$ is PO). 
	We show in the proof that agent 1 cannot yet afford allocation $\allocA$, and so allocation $\allocB$ is in his demand (the rightmost allocation within his budget, using the fact that within his budget, there are no allocations that are also in the blue striped area right of $\allocB$ or in $T_1$). Thus $(\allocB,p)$ is a CE.}
	\label{fig:gamma-case2}
\end{figure}

Now define 
$$
\gamma_i = \max \left\{\frac{b_i}{v_i(\check{\alloc}^i_i)}, \frac{1-b_i}{1-v_i(\check{\alloc}^k_i)}\right\} = \max \left\{\frac{b_i}{b_i^+}, \frac{b_k}{v_i(\check{\alloc}^k_k)}\right\}<1,
$$ 
and note that $\gamma_i$ is well-defined and less than $1$. The assumption that the pair of budgets does not belong to $R_i(v_1,v_2)$ implies that the maximum is obtained by only one of the terms. 
The proof follows by analyzing two cases, as illustrated in Figures \ref{fig:gamma-case1} and \ref{fig:gamma-case2}, respectively:
\begin{figure}[htb]
	\begin{center}
	\includegraphics[width=5in,height=3in]{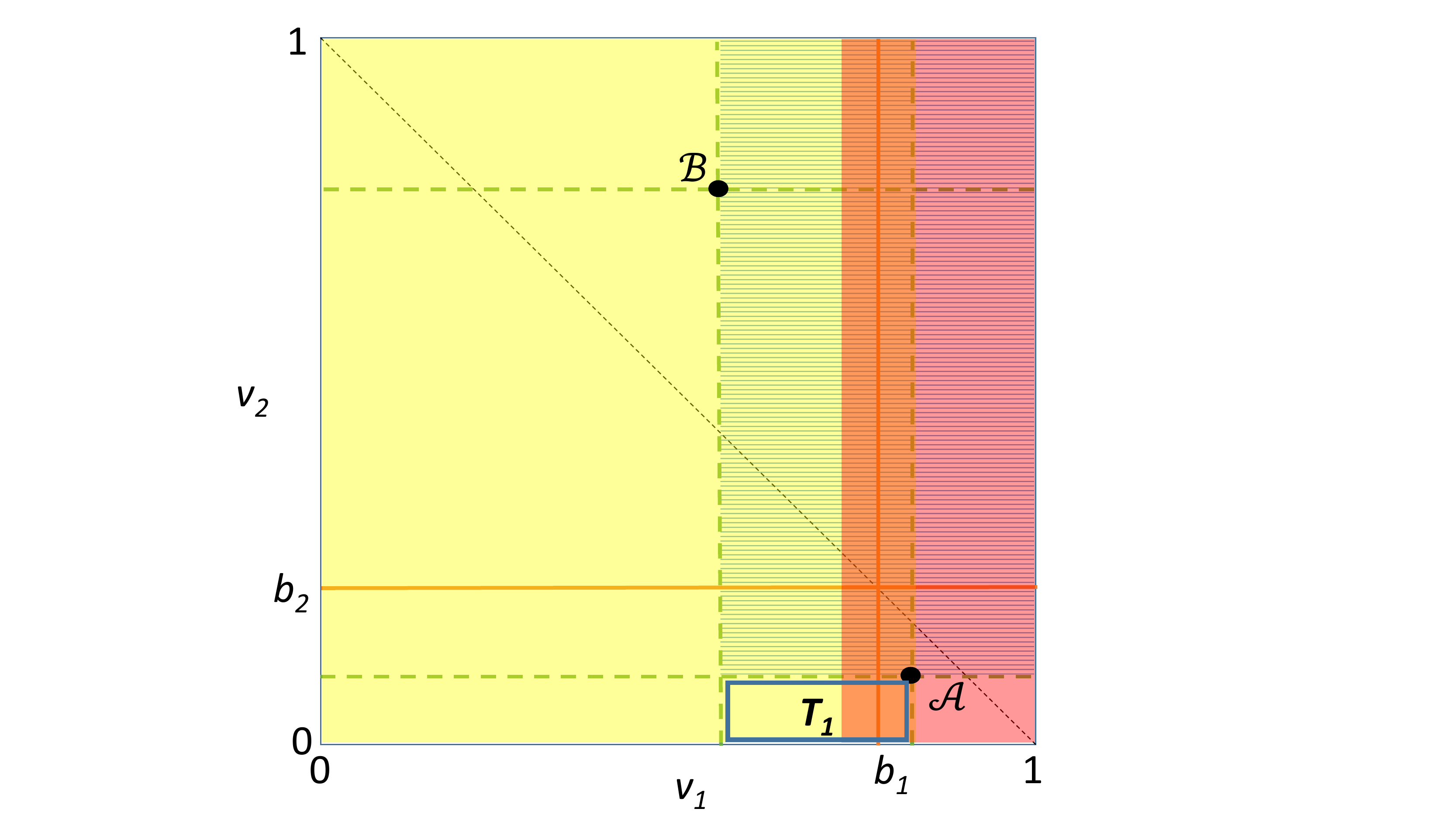}
	\end{center}
	\caption{This figure illustrates Case 1 in the proof of Lemma \ref{lem:2-CE-exist-suff}, using the notation of Figure~\ref{fig:gamma1}, for agent $i=1$. Prices are $p_j=\gamma_1 v_1(\{j\})$ for every item $j$, where $\gamma_1 =b_1/b^1_+$. The solid yellow and red rectangles (at or to the left of $\allocA$, and starting strictly to the left of $b_1$ and stretching to the right, respectively) are the allocations that agents 1 and 2 can afford, respectively. Both agents can afford more allocations than when $\gamma_i=1$ (cf.~Figure \ref{fig:gamma1}). The overlap (depicted as an orange rectangle) contains the allocations that both agents can afford. 
	\newline
	The value of $\gamma_1$ is such that agent 1 can exactly afford allocation $\allocA$, which is clearly demanded by him at these prices (the rightmost allocation within his budget). We show in the proof that agent 2 cannot yet afford allocation $\allocB$ and any other allocation that gives him the same value, and so allocation $\allocA$ is in his demand (the highest allocation within his budget, using that the blue striped area above $\allocA$ that is within his budget, is empty). Thus $(\allocA,p)$ is a CE.}
	\label{fig:gamma-case1}
\end{figure}
\begin{itemize}
	\item {\bf Case 1.} $\gamma_i =b_i/ b_i^+$. We show that $\check{\alloc}^i$ is supported by item prices $p_j= \gamma_i v_i(\{j\})$. For agent $i$, every allocation $\alloc=(\alloc_i,\alloc_k)$ that he can afford satisfies $v_i(\alloc_i)\leq b_i/\gamma_i$, and this holds with equality for $\check{\alloc}^i_i$. For agent $k$, every allocation $\alloc=(\alloc_i,\alloc_k)$ that he can afford satisfies $(b_i/b_i^+) v_i(\alloc_k)\leq 1-b_i$. Since we are in the case that $b_i/b_i^+ > (1-b_i)/(1-v_i(\check{\alloc}^k_i))$, we derive: 
	$$
	\frac{b_i}{b_i^+} (1-v_i(\alloc_i)) =\frac{b_i}{b_i^+} v_i(\alloc_k) \leq 1-b_i < \frac{b_i}{b_i^+} (1-v_i(\check{\alloc}^k_i)), 
	$$
	or equivalently $v_i(\alloc_i)> v_i(\check{\alloc}^k_i)$. We claim that agent $k$'s most preferred allocation that satisfies this is $\check{\alloc}^i$: By Lemma \ref{lem:same-alloc} it holds that $\check{\alloc}^i = \hat{\alloc}^k$, i.e., $\check{\alloc}^i$ is also the PO allocation in which agent $k$ gets at most $b_k$ while maximizing his share. Since there is no allocation $\alloc$ in which $v_i(\alloc_i)> v_i(\check{\alloc}^k_i)$ and  $v_k(\alloc_k)> v_k(\check{\alloc}^i_k)$, the claim follows.
	\item {\bf Case 2.} $\gamma_i = b_k / v_i(\check{\alloc}^k_k)$. We show that $\check{\alloc}^k$ is supported by item prices $p_j= \gamma_i v_i(\{j\})$. For agent $k$, every allocation $\alloc= (\alloc_i,\alloc_k)$ that he can afford satisfies  $v_i(\alloc_k)\leq b_k/\gamma_i$, and this holds as equality for $\check{\alloc}^k_k$.	
	For agent $i$, every allocation $\alloc= (\alloc_i,\alloc_k)$ that he can afford satisfies  $v_i(\alloc_i)\leq b_i/\gamma_i < \frac{b_i}{{b_i}/{v_i(\check{\alloc}^i_i)} } = v_i(\check{\alloc}^i_i) $, as
	$b_i/v_i(\check{\alloc}^i_i)< \frac{1-b_i}{1-v_i(\check{\alloc}^k_i)}=\gamma_i$.  
	Since $T_i$ is empty, it cannot be the case that $v_i(\check{\alloc}^k_i)< v_i(\alloc_i) < v_i(\check{\alloc}^i_i)$. 
	Thus the most preferred allocation that agent $i$ can afford gives him at most $v_i(\check{\alloc}^k_i)$. This is indeed what he gets in allocation $\check{\alloc}^k$, thus $\check{\alloc}^k_i$ is demanded by agent $i$.
\end{itemize} 	
\Halmos
\endproof

\section{Equilibrium Existence for Almost Equal Budgets.}
\label{sec:almost-equal-budgets}
In this section we present our main result for almost equal budgets: the generic existence of a CE for 2 additive agents who are a priori equal. 
The genericity of the budgets serves as a tie-breaking mechanism among the agents, and is sufficient to ensure CE existence for any number of items. The proof utilizes Lemma \ref{lem:2-CE-exist-suff}.
 
\begin{theorem}
	\label{thm:almost-equal}
 	Consider $2$ agents with additive preferences and budgets $b_1>b_2$. 
  	For sufficiently small $\epsilon> 0$, if $b_1-b_2\leq \epsilon$ then there exists a CE that gives every agent his truncated share. 
\end{theorem}

\proof{Proof.}
{\bf Case 1.} Assume first that there exists an allocation which gives each agent a value of exactly $1/2$. Then there is a PO allocation $\alloc$ that gives each agent at least $1/2$. Such an allocation is budget-proportional for $b_1=b_2=1/2$, and thus by Proposition \ref{prop:CE-budget-prop}, there exists a CE $(\alloc,p)$. 
For sufficiently small $\epsilon>0$, let $b_1=(\frac{1}{2}+\epsilon)/(1+\epsilon)>\frac{1}{2}$ and $b_2=1-b_1$ (that is, we slightly increase the budget of agent $1$ while normalizing the sum $b_1+b_2$ to $1$). We claim that $(\alloc,p)$ is also a CE with the perturbed budgets. 
Indeed, as prices have not changed, agent $2$ gets his demand. As for agent $1$, while his budget is slightly larger, he cannot afford any set that is more expensive than his set $\alloc_1$, provided $\epsilon$ is smaller than the difference in prices of any two bundles with non-identical prices.     

{\bf Case 2.} 
Consider now the complementary case, in which no allocation gives each agent a value of exactly $1/2$.
{\bf Case 2(a).} If there is an allocation that gives both agents strictly more than $1/2$, consider any PO allocation that dominates it. 
For sufficiently small $\epsilon$, the PO allocation is budget-proportional for any budgets $b_1>b_2\geq b_1-\epsilon$, and so the result follows from Proposition \ref{prop:CE-budget-prop}.
Note that if there is an allocation that gives both agents strictly less than $1/2$ then the allocation in which the two agents swap their bundles gives both agents more than $1/2$. 

{\bf Case 2(b).} From now on we assume that every allocation gives strictly more than $1/2$ to one agent, and strictly less than $1/2$ to the other agent. For sufficiently small $\epsilon$, such an allocation is neither budget-proportional nor anti-proportional for any budgets $b_1>b_2\geq b_1-\epsilon$.

Recall from Definition \ref{def:trunc-share} that $\hat{\alloc}^1(b_1),\hat{\alloc}^2(b_2)$ are PO allocations that give agents $1,2$ their truncated shares. 
As the set of PO allocations is finite, we can find $\epsilon>0$ such that there is no PO allocation $\alloc$ such that $1/2-2\epsilon<v_1(\alloc_1)<1/2+2\epsilon$. 
For such an $\epsilon$, consider budgets $b_1>b_2\geq b_1-\epsilon$.
Using the notation of Figure \ref{fig:2-points-not-4}, let $\allocA=\hat{\alloc}^2(b_2)$ and $\allocB=\hat{\alloc}^1(b_1)$.

We first claim that $\allocA=\hat{\alloc}^2(1/2)$ and $\allocB=\hat{\alloc}^1(1/2)$.
This holds because $\allocB$ is the PO allocation that gives the largest value to agent~$1$ that is below $1/2-2\epsilon$, but there are no PO allocations that give agent 1 value between $1/2-2\epsilon$ and $1/2$, and thus it also gives the largest value to agent~$1$ that is below $1/2$. A similar argument holds for $\allocA$ and the truncated share of agent 2. 

Because every allocation gives strictly more than $1/2$ to one agent and strictly less to the other,
$v_2(\allocA_2)<1/2\implies v_1(\allocA_1)>1/2$, and $v_1(\allocB_1)<1/2\implies v_2(\allocB_2)>1/2$. 

\begin{figure}
	\begin{center}
	\includegraphics[width=5in,height=3in]{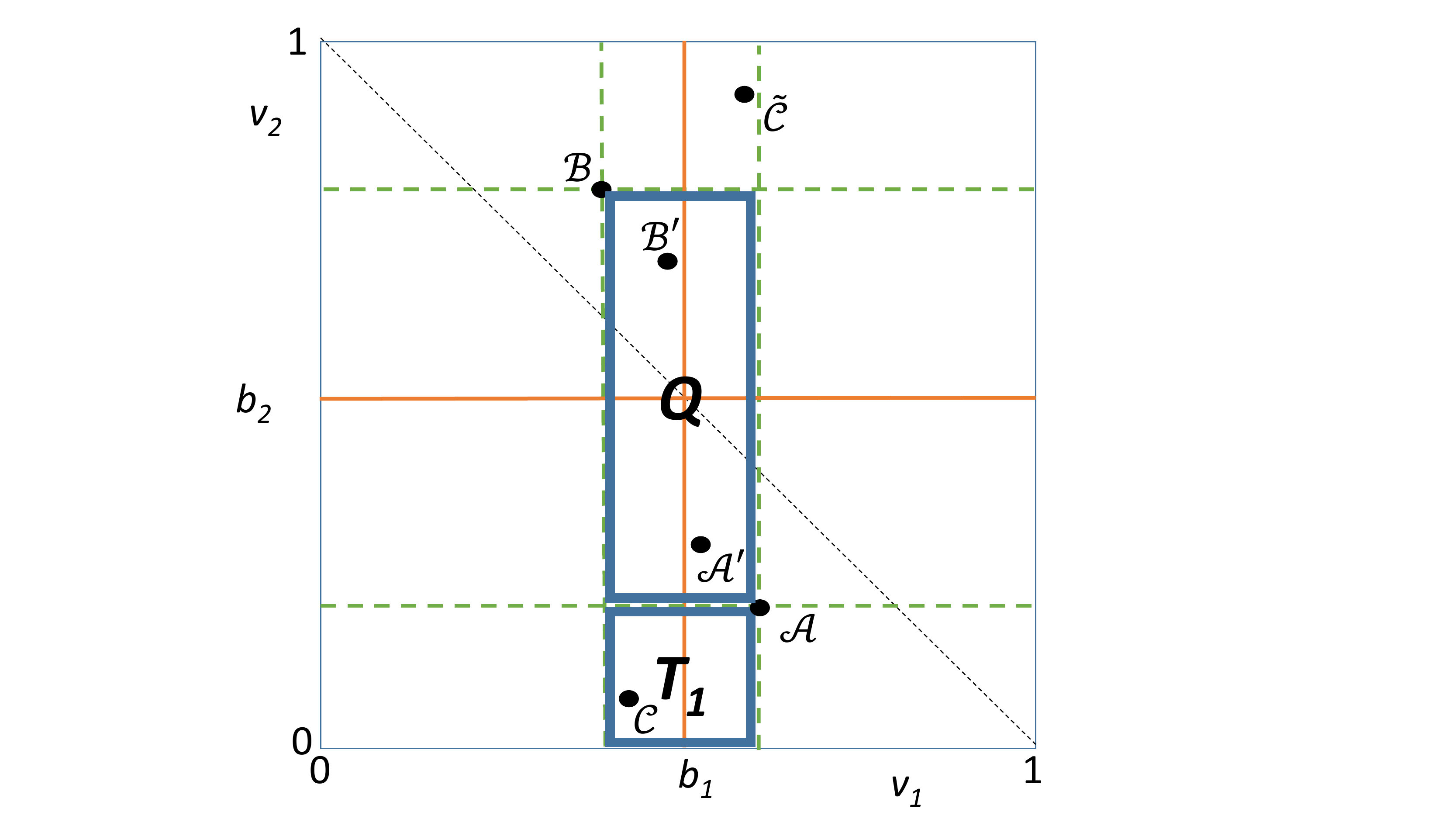}
	\end{center}
	\caption{Illustration of the proof of Theorem \ref{thm:almost-equal}.}
	\label{fig:equal}
\end{figure}

We now show that $\allocA$ and $\allocB$ are ``symmetric'' in the sense that one is obtained from the other by swapping bundles among the agents (and so $v_1(\allocA_1) = v_1(\allocB_2)=1-v_1(\allocB_1)$ and $v_2(\allocA_2) = 1-v_2(\allocB_2)$, as can be seen in Figure~\ref{fig:equal}). 
The proof of the symmetry claim is depicted in Figure~\ref{fig:equal}, which also shows that $T_1,T_2$ (as defined in Definition \ref{def:T-i}) are both empty. 
The budgets $b_1,b_2$ are almost equal, i.e., both are very close to $1/2$. For every allocation $\alloc$ there is a ``symmetric'' allocation $\tilde{\alloc}$ obtained by swapping the allocated bundles, which corresponds to a $180^\circ$ rotation around the point $(1/2,1/2)$. Assume for contradiction that $\allocA$ is not symmetric to $\allocB$, that is $\allocB\neq \tilde{\allocA}=\allocB'$. Then it is also the case that $\allocA\neq \tilde{\allocB}=\allocA'$. Notice that for sufficiently small $\epsilon$, one of $\allocA',\allocB'$ must be located in the interior of the axes-parallel rectangle $Q$ as illustrated in the figure. But since $\allocA=\hat{\alloc}^2(b_2)$ gives agent 2 his truncated share closest to $b_2\approx1/2$ from below, $\allocA'\neq \allocA$ cannot be located as in the figure. Similarly, $\allocB'\neq \allocB$ cannot be located in the interior of $Q$ as in the figure, a contradiction. We have thus established that $\allocA,\allocB$ are symmetric. We now show that the closure of $T_1$ must be empty, except for $\allocA$: If that were not the case -- say, $T_1$ contained an allocation $\allocC\neq \allocA$ -- then its symmetric allocation $\tilde\allocC$ would Pareto dominate $\allocB$ (due to the symmetry of $\allocA,\allocB$), a contradiction. $T_2$ only contains the allocation $\allocB$ by a similar argument using the Pareto optimality of $\allocA$.

We can set $\epsilon$ to be sufficiently small such that $(b_1,b_2)\notin R_i(v_1,v_2)$. 
The proof is complete by invoking Lemma~\ref{lem:2-CE-exist-suff}.
\Halmos
\endproof

\section{Equilibrium Existence for Different Budgets.}
\label{sec:different-budgets}
%
Our techniques for almost-equal budgets apply directly to some cases of different budgets. In this section we derive the generic existence of a CE for 2 additive agents \emph{with the same preferences}, who can be a priori \emph{non}-equal (and in fact quite different) in their entitlement to the items. This is an indivisible version of the \emph{claims (bankruptcy) problem} \cite{Mou02}.

\begin{theorem}
	\label{thm:exist-symm}
	Consider 2 agents with additive preferences and budgets $b_1>b_2$, such that the pair $(b_1,b_2)$ does not belong to the zero-measure subset $R_i$ (Def.~\ref{def:R-i}) for some agent $i$. If the agents have the same preferences then there exists a CE that gives every agent his truncated share. 
\end{theorem}

When both agents share the same additive preferences, we have a ``constant-sum game'' -- whatever one agent gains the other loses. As a consequence, every allocation among such agents is PO, and in addition there is no anti-proportional allocation, in which both agents would get at most their truncated share and one of them strictly so.
Theorem \ref{thm:exist-symm} thus follows directly from the next lemma, whose proof utilizes Lemma \ref{lem:2-CE-exist-suff}:

\begin{lemma}
	\label{lem:all-PO}
	Consider 2 agents with additive preferences and budgets $b_1>b_2$, such that the pair $(b_1,b_2)$ does not belong to the zero-measure subset $R_i$ (Def.~\ref{def:R-i}) for some agent $i$. If every allocation among the agents is PO then there exists a CE. Moreover, if there is no anti-proportional PO allocation then this CE gives each agent his truncated share. 
\end{lemma}

\proof{Proof.}
If there exists a budget-proportional or anti-proportional PO allocation, then there exists a CE by Proposition \ref{prop:CE-budget-prop}. In the former case, this CE clearly gives each agent his truncated share.
Otherwise, the conditions of Lemma \ref{lem:2-CE-exist-suff} hold: any allocation in $T_1$ would be dominated by $\hat{\alloc}^2$, and any allocation in $T_2$ would be dominated by $\hat{\alloc}^1$, but since there are no Pareto dominated allocations these two rectangles must be empty. There thus exists a CE in which every agent gets at least his truncated share, completing the proof. 
\Halmos
\endproof

\section{Conclusions and Discussion.}
\label{sec:discussion}
\subsection{Summary of contributions.}

In this paper we aim to study the solution concept of CE from almost-equal budgets for markets with indivisible items and additive preferences. We focus on simple markets with 2 agents. Even for such markets, the classic solution concept of CEEI is not guaranteed to exist when the indivisibility of the items prevents an allocation that gives each agent her proportional share. 
Our main result shows that a CE from almost-equal budgets -- i.e., equal budgets that are made generic by tiny random perturbations -- \emph{is} guaranteed to exist. The main take-away is that generic almost-equal budgets is a new promising approach for overcoming CE nonexistence. 
The existence results hinge on excluding degenerate market instances,  by adding small noise to the agents' budgets to make them generic; this is in the spirit of smoothed analysis for excluding hard instances to get computational tractability \cite{ST09}. 
We leave as our main open question whether it is possible to apply the approach of generic almost-equal budgets to more than $2$ agents. 

A natural question -- once budgets are non-equal -- is which of the good efficiency and fairness properties of CEEI are maintained by CEs from such budgets?
It is not hard to see that CEs from (possibly very) different budgets are still Pareto efficient (i.e., the first welfare theorem holds). As for fairness, we show that CEs guarantee a new fairness notion: a natural generalization of the $(n+1)$-maximin-share guarantee of Budish. The new notion strengthens the guarantee of Budish for almost-equal budgets, and applies more broadly to arbitrary budgets. It holds in general for CEs with any number of agents and any preferences.

In terms of techniques, our results are based on utilizing the agents' additive preferences as the starting point for equilibrium pricing. 
Our techniques also facilitate two additional results: One is the second welfare theorem for 2 additive agents (complementing the first welfare theorem mentioned above).
The other is a CE existence result for 2 such agents with different budgets and identical preferences, using the same method of budget perturbation.
While we get a lot of mileage from using combinations of preferences for pricing, interestingly this method is insufficient when the 2 agents have arbitrary budgets but non-identical preferences, and we leave this as another intriguing open problem. 

\subsection{Where to go from here.}

\paragraph{Multiple agents.}

The case of $3$ or more agents with almost-equal budgets is qualitatively more complex than $2$ agents due to the following simple observation: If an allocation and budget-exhausting pricing are \emph{not} a CE, then one of the agents wishes to dispose of a subset of her items, and to use the freed-up budget to purchase an alternative bundle of items. When there are more than $2$ agents, the alternative bundle can be owned by \emph{multiple} agents. This means that we cannot expect a characterization like Proposition~\ref{pro:char} to hold for more than $2$ agents. 
In fact we can show that combination pricing is insufficient.
This does not mean of course that generic budgets are not useful for multiple agents with almost-equal budgets. 
One encouraging signal comes from Spliddit instances, in which the preferences are additive and the budgets equal. In all instances with more than 2 agents, after perturbing the budgets to make them almost equal, a CE was found (see Appendix \ref{appx:computer}). Note that an existence result for almost-equal budgets would in particular solve an open problem originating with \cite{PW14} (see also \cite[Section 1.4]{KPW18}), and so we do not expect it to be easy. 
 
One setting in which generic budgets are insufficient to get existence is multiple agents with \emph{different} budgets. In recent work, \citet{Seg18} applies the method of generic budgets we suggest when the number of agents is larger than 2 and the budgets are arbitrary (not almost-equal), and shows that for 4 agents with additive preferences, genericity of the budgets does not guarantee CE existence. 

\paragraph{Fairness among non-equals.}

Fairness notions applicable to the allocation of indivisible items among agents with different entitlements are an under-explored area of theory, despite many real-life applications (e.g., allocating food items among foodbanks catering to different-sized populations \cite{Pre17}), and the increasing importance of endowments as a policy tool (see, e.g., \cite{EMZ18}).
In this paper we develop some of the first such notions, through a classic connection to competitive equilibrium. It is interesting to consider the existence of fair allocations according to our notions independently from the existence of CEs. In particular, an allocation that guarantees every agent her $\ell$-out-of-$d$ share is not necessarily a CE allocation, and can be of independent interest as a solution concept. Another question is what is the appropriate notion of \emph{envy-free} fairness (as opposed to proportionality) for agents with different entitlements
and indivisible items? For a preliminary discussion on this topic, see Appendix~\ref{appx:fairness}.


\section*{Acknowledgments.}
Part of this work was done at Microsoft Research, Herzliya. 

This project has received funding from the European Research Council (ERC) under the European Union's Horizon 2020 research and innovation programme grant agreement No.~740282, and from the European Union's Horizon 2020 research and innovation programme under the Marie Sklodowska-Curie grant agreement No.~708935. 

The authors wish to thank Eric Budish, Herv\'e Moulin, Ariel Procaccia, Assaf Romm, Fedor Sandomirskiy, Erel Segal-Halevi and anonymous reviewers for their very helpful comments.
We thank Segal-Halevi for his simplified proof of Proposition \ref{pro:gen-maximin}, and Procaccia and Shah for kindly providing access to Spliddit data.  
		
\bibliographystyle{abbrvnat}
\bibliography{abb,MOR-bib}

\appendix

\section{CE and Fairness.}
\label{appx:fairness}
\subsection{Detailed fairness preliminaries.}
\label{appx:fair-prelim}

The discussion in this section is summarized in Tables \ref{tab:one} and \ref{tab:two}, which also show where our fairness concepts fit in with some of the existing concepts.

\begin{table}%
	\caption{Fair share and envy-free notions for ordinal preferences (our contribution in bold).}
	\label{tab:one}
	\begin{minipage}{\columnwidth}
		\begin{center}
			\begin{tabular}{|l||l|l|}
				\hline
				& Equal budgets & Arbitrary budgets \\
				\hline
				\hline
				Divisible & Fair share (FS) \cite{Ste48} & Budget-FS \\
				items & Envy-freeness (EF) \cite{Fol67} & Budget-EF%
				\footnote{Budget-FS and budget-EF are natural generalizations of FS and EF, defined in Appendix \ref{appx:fair-prelim}.}
				\\
				\hline
				Indivisible & $1$-out-of-$n$ MMS \cite{Bud11}& {\bf $\ell$-out-of-$d$ MMS} \\
				items&&{\bf (Definition \ref{def:maximin-share})}\\
				& $\EFone$ \cite{Bud11}, $\EFstar$ \cite{CKM+16} & Justified-EF\\
				&&(Definition \ref{def:justified-EF}), open\\
				\hline
			\end{tabular}
		\end{center}
	\end{minipage}
\end{table}%

\begin{table}%
	\caption{Fair share notions for cardinal preferences (our contribution in bold).}
	\label{tab:two}
	\begin{minipage}{\columnwidth}
		\begin{center}
			\begin{tabular}{|l||l|l|}
				\hline
				& Equal budgets & Arbitrary budgets \\
				\hline
				\hline
				Divisible items & Proportionality \cite{Ste48} & Budget-proportionality\\
				&&\cite{RW98,BJK08}\\
				\hline
				Indivisible items & Proportionality \cite{Ste48} & {\bf Truncated share}\\ 
				&&{\bf(Def.~\ref{def:trunc-share})} \\
				\hline
			\end{tabular}
		\end{center}
	\end{minipage}
\end{table}%

\subsubsection{Ordinal preferences.}

An agent $i\in N$ has an \emph{ordinal} preference $\prec_i$ among bundles of items if $S \prec_i T$ for every $T$ preferred to $S$, and there is no representation of the preference via a numerical valuation function $v_i$. 
Fair share and envy-freeness are well-defined for ordinal preferences when items are \emph{divisible}. 
Intuitively, fair share guarantees that each agent believes he receives at least $1/n$ of the ``cake'' being divided, and envy-freeness guarantees he believes no one else receives a better slice than him. More formally,
\emph{fair share} (FS) requires for each agent to receive a bundle that he prefers at least as much as the bundle consisting of a $1/n$-fraction of every item on the market. 
Agent $i$ \emph{envies} agent $k$ given allocation $\alloc$ if $S_i \prec_i S_k$, and an allocation is \emph{envy-free} ($\EF$) if no agent envies another agent.

We observe that for divisible items, both notions extend naturally to budgeted agents: 
\emph{Budget-FS} requires every agent to prefer his bundle at least as much as a $b_i$-fraction of the bundle of all items.
Agent $i$ envies agent $k$ with a larger budget if he prefers a $b_i/b_k$-fraction of $S_k$ to $S_i$, and agent $k'$ with a smaller budget if he prefers $S_{k'}$ to a $b_{k'}/b_i$-fraction of $S_i$; the \emph{budget-EF} property excludes such envy.

When items are indivisible, FS is not well-defined; EF is well-defined but often cannot be satisfied~\citep{DGK+14}.
To circumvent the definition and existence issues stemming from indivisibilities, \citet{Bud11} proposes appropriate variants: 1-out-of-$n$ maximin share (Definition \ref{def:maximin-share}) and envy-free up to one good (Definition \ref{def:EFone}).
\citet{CKM+16} introduce
a strengthening of $\EFone$ called $\EFstar$ (or EFX), envy-freeness up to \emph{any} good, in which for every agents $i,k$ and \emph{any} item $j\in S_k$ it holds that $S_k\setminus \{j\} \prec_i S_i$.

We discuss how to generalize the maximin share guarantee for budgeted agents in Section~\ref{sec:fairness}, defining the notion of $\ell$-out-of-$d$ maximin share (Definition \ref{def:maximin-share}).
We also define a notion called justified-EF for budgets (Definition \ref{def:justified-EF}), and leave the question of how to generalize (non-justified) EF to budgets as an open direction.

\paragraph{Application to CEs.} 
It is well-known that every CE with~$n$ equal budgets gives every agent his 1-out-of-$n$ maximin share and achieves EF. 
\citet{Bud11} shows that every CE with almost equal budgets guarantees 1-out-of-$(n+1)$ maximin share (whereas 1-out-of-$n$ maximin share cannot always be guaranteed). Proposition \ref{pro:gen-maximin} generalizes this to arbitrary budgets using the notion of $\ell$-out-of-$d$ maximin share.

\citet{Bud11} also shows that every CE with almost equal budgets is $\EFone$ (a short proof appears for completeness in Proposition \ref{pro:CE-EFone}). We demonstrate that a CE with almost equal budgets is not necessarily $\EFstar$ (Claim \ref{cla:not-EFstar}), but any CE with arbitrary budgets is justified-EF (Claim \ref{cla:justified-EF}).

\subsubsection{Cardinal preferences.} 
For cardinal preferences, the parallel of FS is the notion of proportionality, which extends naturally to agents with different budgets: Given a budget profile~$b$, an allocation $\alloc$ gives agent $i$ his \emph{budget-proportional share} if agent $i$ receives at least a $b_i$-fraction of his value for all items, that is 
$v_i(\alloc_i)\geq b_i \cdot v_i(M)$.
An allocation is \emph{budget-proportional} (a.k.a.~weighted-proportional) if every agent receives his proportional share. When all budgets are equal, such an allocation is simply called \emph{proportional}. Budget-proportional allocations play a central role in our positive results in Section \ref{sec:proportional}. It is clear that a budget-proportional allocation does not always exist with indivisible items (e.g., with a single one). In Section \ref{sub:truncated} we present a relaxation of budget-proportional share that we call truncated share, which is guaranteed to exist (Definition~\ref{def:trunc-share}).

Budget-EF also extends naturally to agents with cardinal preferences and budgets, by saying that agent~$i$ envies agent $k$ if $v(S_i) < b_i\cdot v_i(S_k) / b_k$. It is not hard to see that budget-EF implies budget-proportionality. 
Another fairness notion for cardinal preferences which naturally extends to budgets (e.g.,~\cite{BGM16}) is \emph{Nash social welfare} maximization. Given a budget profile $b$, an allocation~$\alloc$ is Nash social welfare maximizing if it maximizes $\prod_{i} (v_i(\alloc_i))^{b_i}$, or equivalently, $\sum_i b_i \log v_i(\alloc_i)$, among all allocations (notice that the maximizer is invariant to budget or valuation scaling). 

\paragraph{Application to CEs.} 

In all settings for which we prove CE existence in Sections \ref{sec:almost-equal-budgets} and \ref{sec:different-budgets}, the CEs guarantee every agent his truncated share. 
A CE with equal budgets maximizes the (unweighted) Nash social welfare; in contrast,
we show a simple market with different budgets in which a CE exists, but the allocation that maximizes Nash social welfare is not supported by a CE: 

\begin{claim}
	\label{cla:Nash-social}
	There exists a market of 2 items and 2 agents with additive preferences and unequal budgets, such that the unique PO allocation that maximizes (weighted) Nash social welfare is not supported in a CE, but a CE exists.  
\end{claim}

\proof{Proof.}
Consider 2 agents Alice and Bob with budgets $101,100$, and 2 items $A,B$ valued by Alice $v_1(A)=5, v_1(B)=4$ and by Bob $v_2(A)=1000, v_2(B)=1$.
Clearly the only CE gives item $A$ to Alice and item $B$ to Bob (since Bob cannot prevent Alice from taking item $A$ even if he pays his full budget for it), 
but the only allocation maximizing the Nash social welfare (i.e., maximizing $101\log v_1(\alloc_1) + 100\log v_2(\alloc_2)$) gives item $A$ to Bob and item $B$ to Alice.
\Halmos 
\endproof

\begin{claim}
	\label{obs:anti-proportional}
	There exists a market of 2 items and 2 agents with unequal budgets, for which a CE exists and every CE allocation is anti-proportional.
\end{claim}

\proof{Proof.}
Let $b_1=5/8, b_2=3/8$, $v_1(A)=100$, $v_1(B)=101$, $v_2(A)=1$, $v_2(B)=1000$. Since $b_1>b_2$, in every CE agent $1$ gets his preferred item, item $B$. Moreover, he cannot get both items, as $b_2>b_1/2$ so agent $2$ can always afford at least one item. So in every CE, agent $1$ gets item $B$ and agent $2$ gets item $A$, and their shares are $100/201<5/8$ and $1/1001<3/8$, respectively. Equilibrium prices that support this allocation are $p(A)=3/8,p(B)=5/8$.
\Halmos 
\endproof

\subsection{Variants of envy-freeness.}

In this section we discuss variants of envy-freeness, and define a notion called justified-EF for budgets (Definition \ref{def:justified-EF}). We leave the question of how to generalize (non-justified) EF to budgets as an open direction.

\begin{definition}[EF with indivisibilities \citep{Bud11}]
	\label{def:EFone}
	An allocation $\alloc$ is \emph{envy-free up to one good} ($\EFone$) if for every two agents $i,k$, 
	for \emph{some} item $j\in S_k$ it holds that $S_k\setminus \{j\} \prec_i S_i$.
\end{definition}

\citet{Bud11} establishes an envy-free property of CEs with almost equal budgets, as follows (we include a short proof for completeness):

\begin{proposition}[\citet{Bud11}]
	\label{pro:CE-EFone}
	Consider a CE $(\alloc,p)$ with almost equal budgets $b_1\ge b_2 \ge \dots \ge b_n\ge \frac{m-1}{m}b_1$, then the CE allocation $\mathcal{S}$ is $\EFone$. 
\end{proposition}

\proof{Proof.}
	Fix any two agents $i,k$. We show that there is some item $j^*\in \alloc_k$ such that $\alloc_i \succ_i \alloc_k\setminus \{j^*\}$. 
	By Claim \ref{cla:exhaust-wlog}, we may assume without loss of generality that the budget of agent $k$ is exhausted. So there exists some item $j^*\in \alloc_k$ such that its price $p_{j^*}$ is at least $b_k/|\alloc_k|\geq b_k/m$.
	The price of $\alloc_k\setminus \{j^*\}$ is therefore at most $\frac{m-1}{m}b_k \le b_n\leq b_i$, and so $i$ can afford $\alloc_k\setminus\{j^*\}$. Since $\alloc$ is a CE allocation and bundle $\alloc_k\setminus\{j^*\}$ is within $i$'s budget, $\alloc_i \succ_i \alloc_k\setminus \{j^*\}$ as needed.
	\Halmos 
\endproof

A requirement stronger than (implying) $\EFone$ and weaker than (implied by) $\EF$ is the following: 

\begin{definition}[\citet{CKM+16}, Definition 4.4]
	An allocation $\alloc$ is $\EFstar$ if for every two agents $i$ and $k$, for \emph{every} item $j\in \alloc_k$ it holds that $\alloc_k\setminus \{j\}\prec_i \alloc_i$. 
\end{definition}

An $\EFstar$ allocation always exists for 2 agents -- the cut-and-choose procedure from cake-cutting results in such an allocation. 
It is an open question whether it always exists in general. 
We demonstrate (by an example with non-strict preferences) that an $\EFstar$ allocation is not necessarily $\EF$ even when an $\EF$ allocation exists in the market:

\begin{example}
	Consider $3$ symmetric additive agents, $22$ ``small'' items worth $1$ each, and $2$ ``large'' items worth $7$ each. An $\EF$ allocation is two bundles of 1 large item and 5 small items each, and one bundle of 12 small items. An $\EFstar$ allocation that is not $\EF$ is one bundle of 2 large items, and two bundles of 11 small items each.
\end{example} 

While Proposition \ref{pro:CE-EFone} shows that a CE implies $\EFone$ for almost equal budgets, we next prove that a CE does not imply the stronger property $\EFstar$.

\begin{claim}
	\label{cla:not-EFstar}
	The allocation of a CE from almost equal budgets is not necessarily $\EFstar$, even for~2 symmetric agents with an additive preference over 4 items.
\end{claim}

\proof{Proof.}
	The proof follows from Example \ref{example:CE-no-trunc-share}. 
	Recall that the allocation $(\{A,B\}, \{C,D\})$ is a CE allocation, but it is not $\EFstar$: agent 2 envies agent 1 even if he gives up item $B$. 
	\Halmos 
\endproof

We conclude this section by defining a notion of envy-freeness that a CE with different budgets guarantees.
Borrowing from the matching literature, we define \emph{justified envy} as the envy of an agent with a higher budget towards an agent with a lower budget. The intuition is that any envy of a lower-budget agent towards the allocation of a higher-budget agent isn't justified and so ``doesn't count'', because the higher-budget agent ``deserves'' a better allocation. We thus only care about eliminating justified envy.

\begin{definition}
	\label{def:justified-EF}
	An allocation $\alloc$ is \emph{justified-EF} given budgets $b_1 \ge \dots\ge b_n$ if for every two agents $i<k$, agent $i$ (with the higher budget) does not envy agent $k$ (with the lower budget). An allocation $\alloc$ is justified-EF \emph{for coalitions} if for every agent $i$ and set of agents $K$ such that $i\notin K$ and $b_i\ge \sum_{k\in K} b_k$, agent $i$ does not envy $K$, i.e., $\bigcup_{k\in K}\alloc_k\prec_i \alloc_i$. 
\end{definition}

\begin{claim}
	\label{cla:justified-EF}
	Every CE allocation (with possibly very different budgets) is justified-EF for coalitions.
\end{claim}

\proof{Proof.}
	Assume that $b_i\ge \sum_{k\in K} b_k$. Since the total price $\sum_{k\in K}p(\alloc_k)$ is at most $\sum_{k\in K}b_k$, agent~$i$ can afford the bundle $\bigcup_{k\in K}\alloc_k$. Because $\alloc$ is a CE allocation, it must hold that $\bigcup_{k\in K}\alloc_k\prec_i \alloc_i$.  
	\Halmos 
\endproof

\subsection{Relation to Nash social welfare.}

The first step towards establishing the existence of a market equilibrium is identifying a PO allocation which, given the budgets, can be supported by appropriate prices in a CE.
One natural candidate is the PO allocation that maximizes Nash social welfare -- and thus also enjoys certain fairness properties. 
The next example rules out this approach, by demonstrating the existence of a market with a CE in which the unique PO allocation that maximizes Nash social welfare cannot be supported by prices. 

\begin{example}
Consider 2 agents who value 2 items $A,B$ as follows: $v_1(A)=8$, $v_1(B)=4$, $v_2(A)=1024$, $v_2(B)=1$.
The budgets are $100+\epsilon,100$ where $0<\epsilon\le1$.
The unique CE allocates item $A$ to agent 1 and item $B$ to agent 2, since agent 2 cannot prevent agent 1 from demanding item $A$ -- not even by paying full budget for it. 

The only allocation maximizing the Nash social welfare $(100+\epsilon)\log v_1(\alloc_1) + 100\log v_2(\alloc_2)$ gives item $A$ to agent 2 and item $B$ to agent 1.
\end{example}

\section{Computerized Search for Equilibrium.}
\label{appx:computer}
We attempted to find a market with no CEs for additive as well as general preferences. The instances examined were either randomly generated by sampling from distributions, or taken from real-world Spliddit data. The computational results suggest that CE existence is a wider phenomenon than theoretically verified at this point.

\subsection{Two agents with randomly sampled preferences.}

\paragraph{Setup.}

Our computerized search ran on instances with between 4 and 8 items and 2 agents with randomly generated preferences.%
\footnote{The instances with 4 items were generated as a ``sanity check'', as we know from our companion paper that a CE exists for these instances.} %
We generated both random additive preferences, where the values for the items were drawn from the uniform or Pareto distributions and then normalized to sum up to~1, as well as random general monotone preferences. To generate the monotone preferences we randomly picked an order for all singletons, then randomly placed all pairs among the singletons while maintaining monotonicity, then placed all triplets and so on. 

In choosing budgets for the random additive instances, our goal was to avoid instances for which we know from Lemma \ref{lem:2-CE-exist-suff} or from our companion paper that a CE exists. We thus iterated over consecutive pairs of allocations on the Pareto optimal frontier, and for each such pair tested several budgets that ``crossed'' in between those allocations. To illustrate this, recall Figure \ref{fig:2-points-not-4} in which the budgets ``cross'' between $A$ and $B$. This choice ruled out the existence of budget-proportional allocations. We used additional such considerations to carefully chose the budgets in order to rule out all ``easy cases''. For random non-additive instances, we simply used several choices of arbitrary non-equal budgets.

\paragraph{Running the search.}

For each of the resulting instances we conducted an exhaustive search for an equilibrium: we iterated
over all possible PO allocations, and for each one of them we used CVX with the LP solver MOSEK~7 to look for 
equilibrium prices. 
Note that although the problem is possibly computationally hard, our instances were small enough that they could be completely solved by the solver in a matter of seconds. We verified the equilibria found by the LP solver by implementing a demand oracle. The run time for 10,000 instances of 4 items was several minutes, and run time increased noticeably as the number of items increased.

In all instances with additive preferences that we tested, we found and verified an equilibrium. 
As for general preferences, in all cases with 4 items we found an equilibrium (as expected), and even for 5 items we needed to go over several hundred instances before we found one that does not have an equilibrium. Instances with general preferences that do not have an equilibrium seemed to become more rare as the number of items increased.

\subsection{Spliddit data with additive preferences.}

\paragraph{Setup.}

We ran our second computerized search on instances of Spliddit data, specifically, 803 instances created so far through Spliddit's ``Divide Goods'' application that were kindly provided to us by the Spliddit team \cite[\emph{cf.}][Sec.~4.3]{CKM+16}. In every Spliddit instance, every agent divides a pool of 1000 points among the instance's indivisible items in order to indicate his values for the items; the resulting preference is additive in these values. 

\paragraph{Running the search.}

We implemented a simple t\^atonnement process: Prices start at 0, and all agents are
asked for their demand at these prices. Then the price of over-demanded items is increased  by 1, and the price of 
undemanded items is decreased by 1. Prices thus remain integral throughout the process, and since our budgets are reasonably-sized integers the process is likely to converge 
reasonably quickly (we do not allow it to run for more than $20,000$ iterations). The running time was typically well under a minute, usually no more than a second or two.
One issue that deserves mention (and possibly further research)
is how to update the prices when more than a single item is over- or under-demanded.
Our first attempts either updated only a single such item's price in every iteration, or updated all such items' prices -- both variants converged to a CE fairly often. We improved upon this by
randomly deciding after each price update whether or not to continue updating prices in the current 
iteration. 

\paragraph{Special case of interest.}

An anecdotal but interesting case is non-demo instances with between 5 and 10 items. 
There were 14 such instances available in the data, with between 3 and 9 agents each. 
As Spliddit assumes that agents have equal entitlements, we started by giving all agents equal budgets of 100, in which case an equilibrium was
found for less than half of the instances.
When we added small perturbations to make the budgets only \emph{almost} equal (resulting in the budget vector $(100, 103, 106,\dots)$), we found a CE in all instances.  The same was true for other small perturbations that we tried (resulting in budget vectors like $(100, 101, 104, 109, \dots)$).
We also tried several other budget vectors with budgets that are far from equal (such as $(100, 151, 202, \dots)$ or $(100, 200, 300, \dots)$), and CEs were found for these as well.

\end{document}